\documentclass[final]{siamltex}
\usepackage[dvips]{epsfig}
\usepackage{latexsym}
\usepackage{amsmath}
\usepackage{graphicx}
\usepackage{amssymb}
\usepackage{amsbsy}
\usepackage{subfigure}
\usepackage{wrapfig}
\usepackage{psfrag}

\newcommand{\bal}{{\boldsymbol \alpha}}

\newcommand{\bU}{{\boldsymbol U}}
\newcommand{\be}{{\boldsymbol e}}
\newcommand{\bE}{{\boldsymbol E}}
\newcommand{\bk}{{\boldsymbol k}}

\newcommand{\bd}{{\boldsymbol d}}
\newcommand{\bq}{{\boldsymbol q}}

\newcommand{\bu}{{\boldsymbol u}}

\newcommand{\bx}{{\boldsymbol x}}

\newcommand{\bkap}{{\boldsymbol \kappa}}

\newcommand{\bvep}{{\boldsymbol \varepsilon}}

\DeclareMathOperator{\m}{m}
\DeclareMathOperator{\im}{im}

\newcommand{\Bt}{{\bfm t}}
\newcommand{\Bu}{{\bfm u}}

\newcommand{\CD}{{\mathcal D}}

\newcommand\eq[1] {(\ref{#1})}

\newcommand\pd[2] {{\frac{\partial{#1}}{\partial{#2}}  }}

\newcommand\f[2] {{\frac{#1}{#2} }}

\newcommand{\bfm}[1]{\mbox{\boldmath ${#1}$}}
\newcommand{\nonum}{\nonumber \\}
\newcommand{\beqa}{\begin{eqnarray}}
\newcommand{\eeqa}[1]{\label{#1}\end{eqnarray}}
\newcommand{\beqan}{\begin{eqnarray*}}
\newcommand{\eeqan}{\end{eqnarray*}}
\newcommand{\beq}{\begin{equation}}
\newcommand{\eeq}[1]{\label{#1}\end{equation}}
\newcommand{\bay}{\begin{array}}
\newcommand{\eay}{\end{array}}

\newcommand{\hf}{\frac{1}{2}}

\newcommand{\Tr}{\mbox{ \rm{Tr\,}}}

\newcommand{\If}{\mbox{ \rm{if }}}

\newcommand{\sub}[1]{_{\mbox{\footnotesize{\rm{#1}}}}}

\newcommand{\Ga}{\alpha}
\newcommand{\Gb}{\beta}
\newcommand{\Gd}{\delta}
\newcommand{\Ge}{\epsilon}

\newcommand{\Gg}{\gamma}

\newcommand{\Gl}{\lambda}

\newcommand{\Gm}{\mu}

\newcommand{\Gr}{\rho}

\newcommand{\GG}{\Gamma}
\newcommand{\GL}{\Lambda}

\newcommand{\GO}{\Omega}

\newcommand\rec[1] {\frac{1}{{#1}}}

\newtheorem{remark}[theorem]{Remark}

\numberwithin{equation}{section}
\title{Still states of bistable lattices, compatibility, and phase transition}
\author{
	Andrej Cherkaev
	\thanks{Department of Mathematics, University of Utah ({\tt cherk@math.utah.edu}).
Supported in part by NSF grant DMS-0707974.
}
\and
	Andrei Kouznetsov
	\thanks{Department of Mathematics, Washington State University ({\tt akouznet@math.wsu.edu}).
Work of Andrei Kouznetsov was supported in part by DOE grant DE-FG02-05ER25709.
}
\and
	Alexander Panchenko
	\thanks{Department of Mathematics, Washington State University ({\tt panchenko@math.wsu.edu}). Work of Alexander Panchenko was supported in part by DOE grant DE-FG02-05ER25709 and by NSF grant
OISE-0438765.}
}

\begin{document}
\maketitle
\markboth{A. CHERKAEV, A. KOUZNETSOV AND A.~ PANCHENKO}{STILL STATES AND PHASE TRANSITION}
\begin{abstract}
A two-dimensional bistable lattice is a periodic triangular network of non-linear bi-stable rods. The energy of each rod is piecewise quadratic and has two minima. Consequently, a rod undergoes a reversible phase transition when its elongation reaches a critical value.
We study an energy minimization problem for such lattices. The objective is to characterize the effective energy of the system when the number of nodes in the network approaches infinity.
The most important feature of the effective energy is its "flat bottom". This  means that the effective energy density is zero
for all strains inside a certain three-dimensional set in the strain space. The flat bottom occurs because the microscopic discrete model has a large number of deformed states that carry no forces. We call such deformations \emph{still states}.  In the paper, we present a complete characterization of the "flat bottom" set in terms of the parameters of the network.  This is done by constructing a family of still states whose average strains densely fill the set in question.

The two-dimensional case is more difficult than the previously studied case of one-dimensional chains, because the elongations in two-dimensional  networks must satisfy certain compatibility conditions that do not arise in the one-dimensional case.
We derive these conditions for small and arbitrary deformations. For small deformations a complete analysis is provided.
\end{abstract}



\section{Introduction}

This paper suggests a simple discrete model for phase transition
theory.  In two dimensions, we study deformation of {\it bi-stable lattices} (BL), which are periodic triangular networks of bistable
elastic rods. A similar network with linearly elastic rods has been
used by Cauchy,
who formally averaged this assembly and obtained one of the first equations of
continuum elasticity \cite{theory_elasticity_love}. Here, we
consider the same assembly, assuming, however, that the rods are
bi-stable. The bi-stability models phase transitions: it is assumed
that each rod has two equilibrium states that of length
$l$ and $l (1+s)$, respectively. Here, $s$ can vary between $\hf$ and $2$, so that
a triangle of transformed rods still remains a triangle.

We believe that this bi-stable lattice model
is the simplest and natural finite dimensional model
for phase transitions in solids. Although it does not have all the
features of continuum model, it appears to capture most of the
expected features. At the same time this model allows for detailed
description of inhomogeneous deformations and does not involve ad
hoc assumptions of the continuum theory.

One dimensional chains of bi-stable rods are well investigated
\cite{braides_interactions, braidis, cherkaev_1, cherkaev_2,  truskinovsky-siam, truskinovsky}.
Each bistable element can be in one of the two  states ({\em short}
and {\em long} mode) that differ by the equilibrium length. It is
assumed for definiteness that initially all rods are in short mode
and can transit to the long mode.  The state of a chain with some
transited elements is characterized by a scalar relative elongation, or one-dimensional strain, that is
the normalized difference between the length in the deformed configuration
and reference length.

Some results on the two-dimensional bi-stable lattices can be found in \cite{slepyan_surprise, slepyan04-waves} (explicit construction of certain special solutions, discrete Green functions), and \cite{cherkaev_vinogradov, cherkaev_protective} (direct numerical simulations of time-dependent impact fracture problems). Other related problems are treated in the book \cite{slepyan-book} that also contains a wealth of additional references.

The present work differs from the above mentioned papers both in approach and in scope.
It addresses the specific difficultly of the two-dimensional
problem --  \emph{compatibility
conditions} on the rods' length.  These conditions are automatically satisfied in
one dimensional chains. In two-dimensional lattice, the transition of links to a new state
generally leads to elastic deformation of all rods. For example, one
transformed (elongated) rod does not fit into a triangular lattice
of non-transformed (short) links without distorting the lattice.
Each such distortion increases the energy of the assembly.

One of the objectives of the paper is to describe special lattice deformations that we call
{\em still states}. These deformations carry zero forces in each rod, and thus the energy of each still state is zero.  In any such state the length of each rod should be either $l$ or $l(1+s)$.
Since the number of rods is larger than the number of nodes times the dimension, an arbitrary assignment of lengths may not correspond to an actual lattice deformation. Therefore
compatibility conditions must be satisfied at a still state.

We describe the set of
average strains of still states. First, we consider the network of rods undergoing arbitrary
deformations. We assume that  an effective strain tensor is represented by its
eigenvalues. In the plane of eigenvalues, the set ${\mathcal H}_L$ of all still
states is a nonconvex curvilinear hexagon. We demonstrate structures
that densely fill ${\mathcal H}_L$ and derive nonlinear compatibility
conditions.

Further, we consider the case of small deformation, linearize the
compatibility conditions and develop  the theory in more detail.
We show that the linearized compatibility conditions are necessary and sufficient for existence of a
lattice deformation with a given set of link elongations. In this
sense, the conditions are analogous to the compatibility
conditions in continuum elasticity. We also describe a sequence of still
states whose average strains reach the average strain of any still state by incremental change of the
structure starting from the reference configuration. This
procedure can describe an evolutionary irreversible process of phase
transition, such as the damage distribution. In this case, the
initial state corresponds to  "undamaged" configuration and structural change -- to irreversible
damage propagation.

We also construct the set $\mathcal{D}$ of all average strain
tensors corresponding to still states and sets of deformed elements that densely fill that
set. In the space of entries of the strain tensor,  the
set $\mathcal{D}$  turns out to be a parallelepiped. We obtain explicit formulas that relate the side lengths and orientation of this parallelepiped to the non-dimensional critical elongation parameter $s$, and three lattice direction vectors.

Finally, we suggest an equation for the effective energy density as a function of the total strain.
It
is as follows: if a strain is inside of the set
$\mathcal{D}$, then the energy density is asymptotically close to zero as
the number of the network nodes goes to infinity. If the strain is
outside of this set, the energy density is proportional to the
square of the distance between the strain tensor and
$\mathcal{D}$.

The paper is organized as follows. Section 2 contains formulation of the problem and derivation of compatibility conditions
for arbitrary deformations. Small deformation compatibility conditions are also obtained here. In Section 3 we describe several explicit constructions of still states for arbitrary deformations, as well as a set of eigenstrains of all still states. The remainder of the paper deals with the small deformation case. In Section 4.1 we formulate the problem. A new formula for average strain is obtained in Section 4.2. Unlike existing formulas that relate average strain to nodal displacements, our formula gives an expression in terms of rod elongations. This allows to relate components of strain to the average elongations along lattice directions. Such formulas seem to be of independent interest.  In section 5 we provide a complete analysis of the compatibility conditions.
In Section 6, we prove a characterization theorem for ${\mathcal D}$. This is done by constructing another family of still states called stripes. The eigenstrains of the stripes fill ${\mathcal D}$ densely, in the sense that every strain within this set can be approximated by an average strain of some stripe, with an error of order $N^{-1/2}$ where
 $N$ is the number of nodes.  Section 7 contains a discussion of the effective energy density. Conclusions are provided in Section 8.


\section{The problem} \label{section:problem}
\subsection{Energy of one bistable rod} \label{subsection:one_dimensional_rod}
Consider a  bistable elastic rod with the energy $W_r(x)$ that
possesses two equal minima $ W_r(l)= W_r(\alpha l)=0$. Let us also
assume that $W_r(x)$ is convex outside of the interval $[l, ~l(1+s)]$,
and its
second derivative is positive in a proximity of each minima $x=l$ and
$x=l(1+s)$.
For definiteness, we may assume that the dependence of $W_r(x)$ on the
length $x>0$
is piece-wise quadratic $(W_r^q)$ or polynomial $(W_r^p)$
\begin{eqnarray}
\label{rod-energy} W_r^q(x) &=& \min\left\{ \hf (x-l)^2, \hf C
(x-l(1+s))^2\right\}, \nonum
W_r^p(x) &= &  (x-l)^2
(x-l(1+s))^2,
\end{eqnarray}
where  $l$ is the length of the rod in the reference configuration,
$l+s$ is the length in the elongated mode, and
$\hf <1+s< 2$.

The rods are  elastic and locally stable in a proximity of the
equilibria. The elastic force $f_r$ in the rod is
 $$
f_r(x)=\pd{W_r}{x}.
 $$
The rod has two equilibrium states $x=l$ and $x=l(1+s)$ of equal
energy
$$ f_r(x)=0, \quad  W_r(x)=0 \quad \If x=l,\;l(1+s).$$
The magnitude of the force monotonically increases with the elongation $l$ in the proximity of equilibria.

There are several equilibrium lengths $x_\Ga$ and $x_\Gb$ for every
$f_r$ in a proximity of zero. They are solutions of the equation
$$ \pd{W_r}{x}= C $$

\begin{figure}
        \begin{center}
                \psfrag{w}{$W_r(x)$} \psfrag{l(1+s)}{$l(1+s)$} \psfrag{x}{$x$} \psfrag{l}{$l$}
                \includegraphics[width=3in]{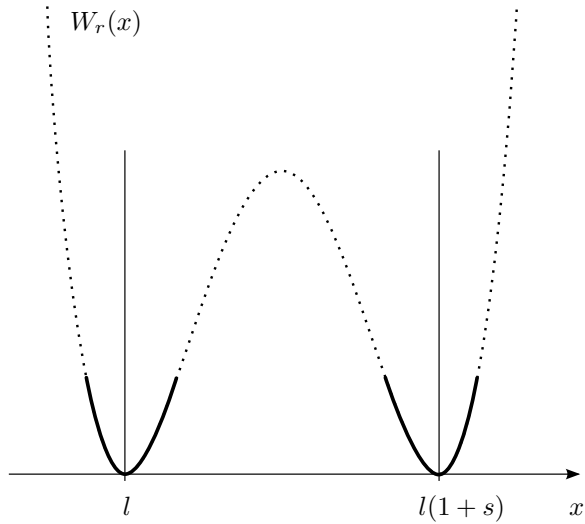}
                \caption{Energy of a bistable rod} \label{f:1}
        \end{center}
\end{figure}

\subsection{Triangular network}

Consider a triangular periodic network made of the links defined above. For each node, there are
six rods joining it with six neighbors.
When the rods are of equal lengths, the network is elastically
isotropic, its Poisson coefficient is equal to $\rec{4}$,
\cite{theory_elasticity_love}. The number of nodes in the period is three times less
than the number of rods between them. There are three families of the
parallel rods in the network

\begin{figure}
\begin{center}
\includegraphics[height=2in]{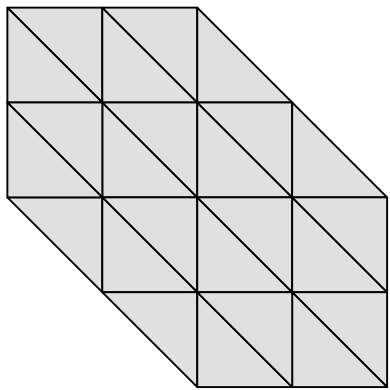}
\caption{To compatibility} \label{f:2}
\end{center}
\end{figure}

When the rods transit to a different state, the network becomes
inhomogeneous (each rod can have a different length). We assume that the transition is periodic. Each period
consists of $N$ nodes where $N$ can be arbitrary large. The
network's energy is the sum of energies stored in all $3 N$ links.
The length of a link can be expressed through the position of its ends, that
is the nodes. The nodes are determined by $N$ pairs of coordinates
in a plane, or by $2N$ parameters. Hence, the links' lengths can not be
arbitrary (for example, see Figure \ref{f:2}). We conclude that they
are subject to $N$ {\em compatibility conditions} that we
derive now.

\subsection{Compatibility conditions}

\subsubsection{Necessary condition}
There are conditions that constrain the lengths of the rods that can
be joined in a hexagonal lattice. Consider an inner node in the
lattice. There are six rods that link the node with the neighbors;
let us denote the lengths of them as $a_1, \ldots a_6$. Consider
also six rods that surround the node forming a hexagon around it;
let us denote the lengths of them as $b_1, \ldots b_6$. These rods
form six triangles joined at the node. The angles at the node will be denoted by $\phi_1, \ldots,
\phi_6$ .

The lengths of the listed twelve rods cannot be arbitrary.  The sum
of angles of the six triangles $a_1a_2, b_1$, .. is equal to $2
\pi$,
 \beq
\sum_{i=1}^6 \phi_i=2\pi.
 \eeq{comp1}
To express this condition in terms of the lengths, we use
trigonometry. Cosine of each angle $\phi_i$ is expressed through the
lengths of the links as
 \beq
\cos( \phi_i)=\frac{a_i^2 + a_{i+1}^2  -  b_i^2}{2 a_i a_{i+1}},
\quad i=1, \ldots, 6.
 \eeq{comp2}
Here, $a_{7}=a_1$. Therefore, the lengths of rods are constrained as
follows \beq \sum_{i=1}^6 \arccos \left(  \frac{a_i^2+ a_{i+1}^2   -
b_i^2}{2 a_i a_{i+1}}\right) =2\pi. \eeq{comp3} The number of these
constraints is equal to the number of inner nodes in the structure.

\subsubsection{Linearized compatibility conditions} In a linearized case when the
lengths of the rods are close to $l$, constraints \eq{comp3} are
simplified. Assuming that
 \beq
a_i=l(1+ \kappa_{a_i}), \quad b_i=l (1+ \kappa_{b_i}), \quad i=1, \ldots, 6
 \eeq{comp4}
we write $\phi_i$ as the function of $\kappa_{a_i}$, $\kappa_{a_{i+1}}$, $\kappa_{b_i}$:
$$ \phi_i = \arccos \frac{(1+ \kappa_{a_i})^2 + (1+ \kappa_{a_{i+1}})^2  -  (1+ \kappa_{b_i})^2}{2 (1+ \kappa_{a_i}) (1+ \kappa_{a_{i+1}})} . $$
Linearizing near
$$\left(\begin{array}{l}
\kappa_{a_i} \\
\kappa_{a_{i+1}} \\
\kappa_{b_i}
\end{array}\right) = \left(\begin{array}{c}
0 \\
0 \\
0
\end{array}\right) $$
we find
$$ \sum_{i=1}^6 \phi_i = 2 \pi + \frac{\sqrt{3}}{3}\left(\sum_{i=1}^6 \kappa_{a_i} + \sum_{i=1}^6 \kappa_{a_{i+1}} - 2 \sum_{i=1}^6\kappa_{b_i}\right) $$
Substituting this expression into \eq{comp1}, we obtain an elegant
{\em linearized compatibility condition for lattices}
\beq
\sum_{i=1}^6 \kappa_{a_i} = \sum_{i=1}^6\kappa_{b_i}
\eeq{comp6}
It states that the sum of the elongations of the spokes that come out of a
node is equal to the elongation of the hexagonal rim around this node.


\section{Eigenstrains and still states}
\subsection{Definitions}
\subsubsection{Still configurations}
In this section we consider {\em still configurations}. These are deformed states with zero force
in each rod. Because of the force-elongation dependence, this requires the length of each rod to be either
$l$ or $l(1+s)$. To simplify presentation, in this section we scale $l=1$ and denote $l(1+s)=a$.
Given a collection of such rod lengths, we call a
configuration {\em still} if it is geometrically compatible, that is
the ends of the link of the length one and $a$ can meet in the
nodes. The examples of still configurations are given below in
Figure \ref{fig:hex_strips}. This section contains somewhat informal, but explicit description of the set of
(average) {\em eigenstrains} of all still states corresponding to arbitrary (not necessarily small), deformations. We recall that eigenstrain is a generic name given to various non-elastic strains, such as strains due to thermal expansion, phase transformation, initial strains and so on.

Notice that the problem is reduced to a geometric problem of {\em
tiling} of the plane with triangles of four types with the
lengths of the sides being equal to $(1, 1, 1)$, $(1, 1, a)$, $(1,
a, a)$. and $(a,a, a)$, respectively.

Any still configuration can be split into parts that are also still
configurations. Each configuration is characterized by a symmetric
second-rank {\em eigenstrain tensor } which shows the change
of the initial shape of a periodic cell. Assume that a unit square
element of periodicity of an initial (non-transformed) network
contains a very large number of triangles. After deformation, it
changes its shape. Asymptotically, this element is close to a
parallelogram. The eigenstrain tensor describes the shape of
it: the lengths of its sides and the angle between them.

\subsubsection{Eigenstrain of still states}
Generally, a deformation due to transformation is followed by
elastic deformation of all links due to compatibility conditions
\eq{comp3}. Therefore, eigenstrain of the composite network is
not a convex combination of eigenstrains of its components that
are, in principle, incompatible and lead to additional distortions
on the boundaries of components.
 The eigenstrain of a compatible
configurations of still states is an affine function (the average)
of the eigenstrains of the phases. It is independent of the
structure of the mixture
 \beq
\bE\sub{mixture}=\sum_i^k\Gm_i \bE_i,  \quad \sum_i^k\Gm_i=1 , \quad
\Gm_i =\f{k_i}{N}, ~i=0, \ldots, N
 \eeq{10}
where $\mu_i $ is the volume fraction of the phase with
eigenstrain $\bE_i$. The only role of the structure is the selection
of compatible states.

Next, we describe the variety of still states.
\subsection{The homogeneous still states}
In the initial network, there are three families of codirected rods.
The {\em homogeneous still states} are the states in which all the
rods in one family are in either "short" or "long" mode.
\subsubsection{Classification of homogeneous states}
There are eight $(2^3=8) $ homogeneous states that consist of
periodic arrays of equal triangles and their reverses.

\begin{figure}
\begin{center}
\includegraphics[height=3in]{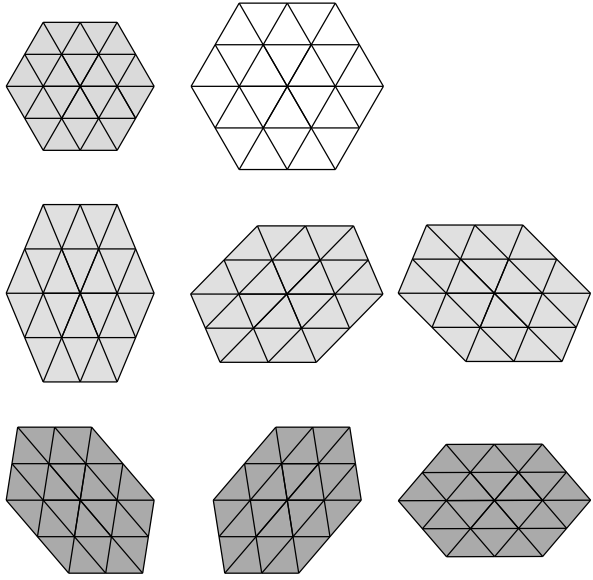}
\caption{The homogeneous still states} \label{f:3}
\end{center}
\end{figure}

\begin{enumerate}
\item
Initial state $S_\Ga$ with the lengths of the links equal to $(1, 1,
1)$. The elementary square of the state (the eigenstrain) is
assumed to be an identity matrix,
 \beq
\bE_\Ga=\begin{pmatrix} 1 & 0 \cr 0 & 1
\end{pmatrix}
 \eeq{bas1}
\item
The dual state $S_\Gb$ with the lengths of the links equal to
$(a,a,a)$. The unit square element $E_\Ga$ of the initial state is
transformed to
 \beq
\bE_\Gb=\begin{pmatrix}a & 0 \cr 0 & a
\end{pmatrix}
 \eeq{bas2}
and it is still proportional to an identity matrix. We call $\bE_\Gb$
the eigenstrain of the state $S_\Gb$.
\item
State $S_\Gg$ with the lengths of the links equal to $(a,a, 1)$ that
consists of  isosceles triangles. The  square element $\bE_\Ga$ of the
initial state is transformed to
 \beq
\bE_\Gg=\begin{pmatrix}\f{\tan  \Gg}{\sqrt{3}} & 0 \cr 0 & 1
\end{pmatrix}
 \eeq{bas3}
Here, $ \Gg$ is the angle between the base of the isosceles triangle
and its side, $\Gg$ is expressed through $a$ as
$$
a=\rec{2 \cos \Gg}
$$
 $\bE_\Gg$ is the eigenstrain of the state $S_\Gg$.

\item Together with $S_\Gg$, there are two twin states, $S_\Gg'$ with
the lengths of the links equal to $(a,1, a)$ and $S_\Gg''$ with the
lengths of the links equal to $(1,a, a)$. The eigenstrains of
these states are obtained from $\bE_\Gg$ by rotating it to the angle
of $\pm 60^\circ$, respectively.
 \beq
\bE_\Gg'=R^T\,\bE_\Gg\,R \quad \bE_\Gg''=R\,\bE_\Gg\,R^T
 \eeq{tw1}
where $R$ is the  $60^0$ rotation matrix.
\item
State $S_\Gd$ with the lengths of the links equal to $(1,1,a)$ that
consists of  isosceles triangles. The  square element $\bE_\Ga$ of the
initial state is transformed to
 \beq
\bE_\Gd=\begin{pmatrix}\f{2\sin \Gd}{\sqrt{3}} & 0 \cr 0 &
a
\end{pmatrix}
=
\begin{pmatrix}\f{2\sin  \Gd}{\sqrt{3}} & 0 \cr 0 & {2 \cos \Gd}
\end{pmatrix}
 \eeq{bas4}
Here, $ \Gd$ is the angle between the base of the isosceles triangle
and its side, $\Gd$ is expressed through $a$ as
$$
a={2 \cos \Gd}
$$
The angles $\Gd$ and $\Gg$ are related as
$$
\cos \Gg\,\cos \Gd=\rec{4}
$$
\item
State $S_\Gd$ is also accompanied by the two twin states,
$S_\Gd'$ with the lengths of the links equal to $(a,1, 1)$ and
$S_\Gd''$ with the lengths of the links equal to $(1,a, 1)$. The
eigenstrains of these states are obtained from $\bE_\Gg$ by
rotating it by the angle of $60^\circ$, and $-60^\circ$, respectively.
 \beq
\bE_\Gd'=R^T\,\bE_\Gd\,R \quad \bE_\Gd''=R\,\bE_\Gd\,R^T
 \eeq{tw2}

\end{enumerate}

\begin{remark}
Pairs $S_\Ga$ , $S_\Gb$ and $S_\Gg$ , $S_\Gd$ are dual: each of them
is transferred into another by mutually replacing rods of the length
$1$ and $a$.
\end{remark}

\subsubsection{Vertices of the eigenstrain states set}
Consider a set $\CD$ of all possible eigenstrains. In
a three-dimensional space where coordinates are entries of the eigenstrain tensor, this set is a
bounded region. The eigenstrains of the homogeneous states correspond
to the vertices of $\CD$. Indeed, there the elongation of the each family of the rods is extreme, that is the deformation
in the directions of the rods' families is extreme. The rods cannot be elongated
(contracted) in these directions without an elastic deformation.
Therefore, there exist three trial matrices
$\bE_{{tr}_1}$ $\bE_{{tr}_2}$ and $\bE_{{tr}_2}$ of deformation that correspond to impossible change of
eigenstrain in the sense that
$$\bE_z+ \Ge \bE_{{tr}_i}, \quad z=\Ga,\ldots, \Gb, ~i=1,2,3 $$
does not correspond to any still state, for any $\Ge>0$.

Consider, for example, state $S_\Ga$. Set of impossible deformations
$\bE_{tr}$ contains a cone of matrices  of the form
\beqa
	\bE^\Ga_{tr} = \sum_{i=1}^3 d_i\Bt_i\Bt_i^T \label{trial-a}\\
	d_i<0,
	\quad \Bt_1 = \begin{pmatrix} 1\cr 0 \end{pmatrix} ,
	\Bt_3 = \hf \begin{pmatrix} 1 \cr \sqrt{3} \end{pmatrix},
	\Bt_3 = \hf \begin{pmatrix}-1 \cr \sqrt{3} \end{pmatrix},\nonumber
\eeqa
where $d<0$ is arbitrary.
These deformations require contraction of one of the rods. In the state $S_\Ga$ such contraction is
impossible. The impossible deformations form a cone in
the eigenstrain space. Similar cones can be established for all
homogeneous states because the structure cannot be extended
(compressed) in the direction  of  long (short) rods.
Below we describe the two- and three-phase composites of still states.
The eigenstrain of any still state composite is a convex combination of eigenstrains of its phases,
therefore, the two-phase composites correspond to edges of $\CD$ and  two-phase composites -- to its phases.
Set $\CD$ is a polyhedron.

We also describe a two-dimensional set ${\mathcal H}_L$ of eigenvalues $(\GL_A, \GL_B)$ of eigenstrains.
This set is also bounded and symmetric (includes pairs $(\GL_A, \GL_B)$ and $(\GL_B, \GL_A)$).
The eigenstrains of the homogeneous states correspond
to vertices of ${\mathcal H}_L$, and the differently oriented homogeneous still states $S_\Gg$, $S_\Gg' $ and $S_\Gg''$ (and
$S_\Gd$, $S_\Gd' $ and $S_\Gd''$) correspond to the same pairs of eigenvalues -- vertices of ${\mathcal H}_L$.
However, the eigenvalues of a two-phase composite are convex combinations of eigenvalues of components only if
their vertices are coaxial tensors (the eigen-axes are codirected). Otherwise, they are located along a curve that
join the vertices. Therefore,  ${\mathcal H}_L$ is a symmetric curvilinear hexagon.

\subsection{Laminate structures}
There are still states composed from several basis homogeneous
states. We call two still states $\bE_y$ and $\bE_z$ are called {\em laminate
compatible} if lengths of their rods along the layers are the same.
In terms of eigenstrains, the states are laminate compatible if
a vector $\tau$ (the tangent) exists such that
 \beq
\tau^T(\bE_y-\bE_z) \tau=0
 \eeq{comp11}
that is, matrices $\bE_y$ and $\bE_z$ have the same tangent component. All
pairs of homogeneous states except $S_\Ga$ and $S_\Gb$ can be
laminated together, along the side of a common self-strain, if the states are allowed to rotate freely.
We also remark that threads in laminate-compatible structures may possess kinks, as in Figure 4. The continuity of the angles is not required for compatibility of a lattice, in contrast with continuum mechanics.

\subsubsection{Structures from an isotropic and anisotropic phases}
States $S_\Ga$ and $S_\Gd$ are laminate compatible. The
eigenstrains of mixture are
$$\bE_{\Ga\Gd}=\mu \bE_\Ga+ (1-\mu) \bE_\Gd, ~~~~~~~~\mu=\f{k_\Ga}{k_\Ga+k_\Gd}$$
where $\mu$ is the volume fraction of phase
$S_\Ga$, $k_\Ga$ is the thickness of the layer,  and $k_\Gd$ is the
thickness of the layer of phase $S_\Gd$. The laminate is shown in
Figure \ref{fig:hex_strips}.

In addition, the eigenstrains $\bE_\Ga$ and $\bE_\Gb$ are isotropic
and therefore their eigenvectors are codirected with eigenvalues of
any phase they are mixed with. A structure of one of them and
another phase leads to average of eigenvalues $\Gl_1, \Gl_2$ of
eigenstrain. The eigenvalues $\Gl^x_{\Ga \Gd}, \Gl^y_{\Ga \Gd}$
of $E_{\Ga \Gd} $ are located at the straight line between
eigenvalues of $\bE_\Ga$ and $\bE_\Gd$.
$$
\Gl^x_{\Ga \Gd}=\mu \Gl^x_{\Ga}+ (1-\mu) \Gl^x_{\Gd},
\quad \Gl^y_{\Ga \Gd} =\mu \Gl^y_{\Ga}+ (1-\mu) \Gl^y_{\Gd}
$$
Similar laminates can be built for $\bE_{\Ga\Gg}$, $\bE_{\Gb\Gd}$,
$\bE_{\Gg\Gb}$, and $\bE_{\Gd\Gg}$. The eigenvalues of these structures
span the intervals between the eigenvalues of the homogeneous states.

\begin{figure}
	\begin{center}
		\subfigure[an isotropic and anisotropic phases]{
			\label{fig:laminates_ai}
			\includegraphics[width=0.25\textwidth]{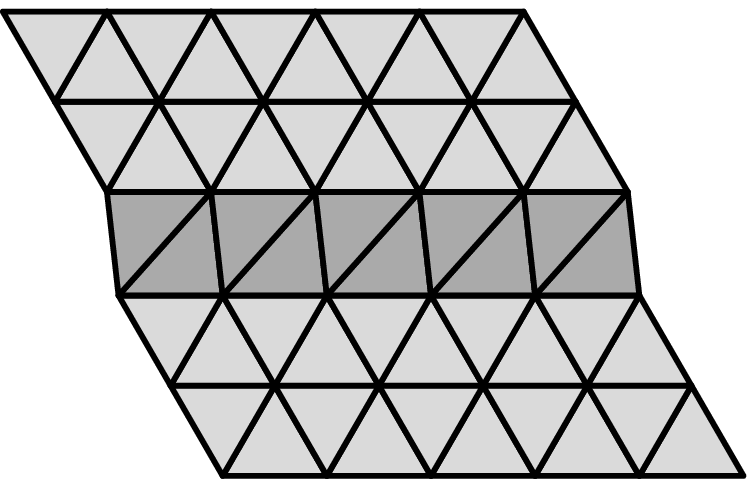}
		}
		\subfigure[two-phase twin]{
			\label{fig:laminates_tw}
			\includegraphics[width=0.25\textwidth]{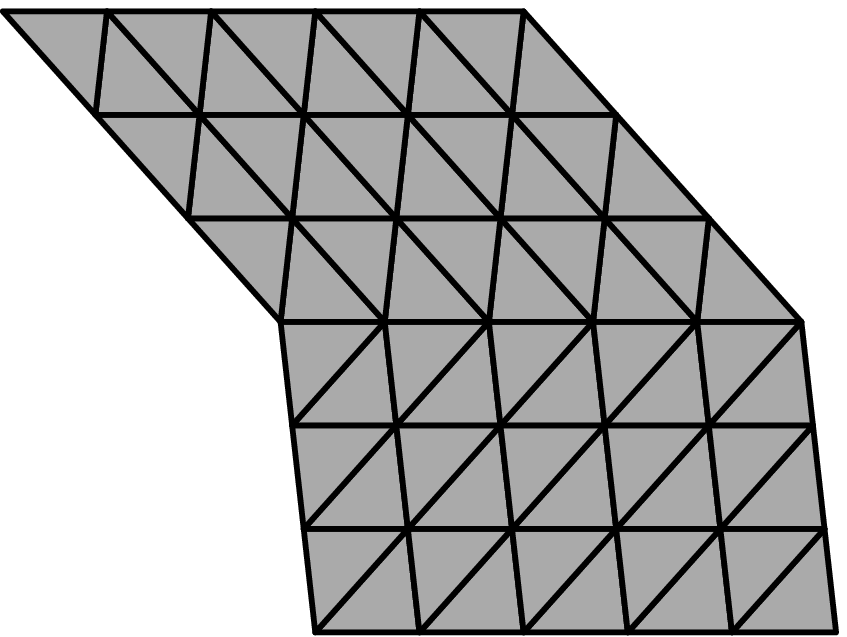}
		}
		\subfigure[two anisotropic phases]{
			\label{fig:laminates_aa}
			\includegraphics[width=0.25\textwidth]{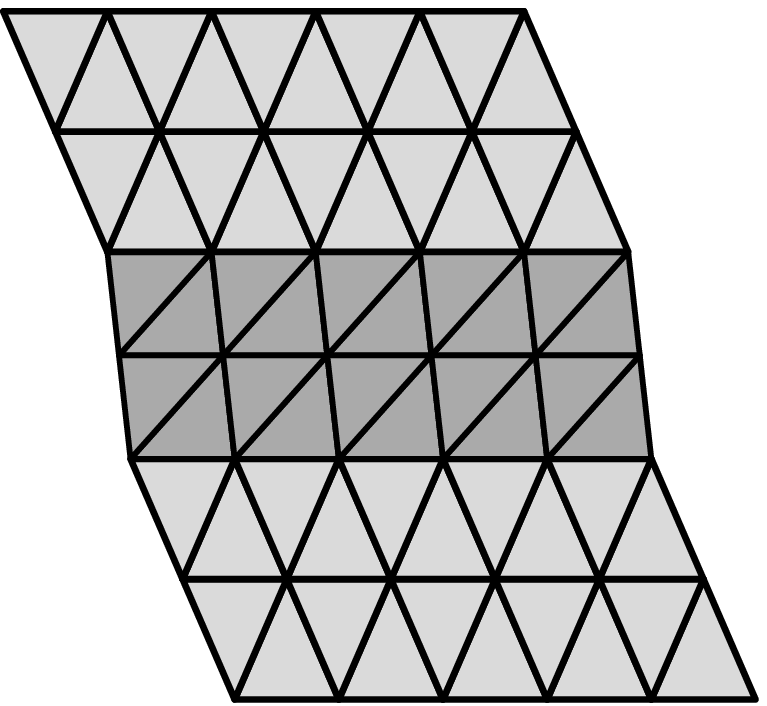}
		}
	\end{center}
	\caption{Examples of laminates}
	\label{fig:4}
\end{figure}

\subsubsection{Laminates from two anisotropic phases $S_\Gd$ and $S_\Gg$}
Consider the compatibility of two anisotropic states  $S_\Gd$.
Each of them consists of "long" and "short" rods.
Assume for definiteness, that tangent $\tau$ to laminates corresponds to the line of  "short" links.
This line is an axis of symmetry for the state $\bE_\Gg$ that consists of two families of long
and one family of short rods. In orthogonal coordinates $\tau, \nu$, the eigenstrain $\bE_\Gg$
is diagonal, its eigenvectors are codirected with $\tau$ and $ \nu$.

The state $\bE_\Gd$ (that consists of two families of short
and one family of long rods) can be made compatible with $\bE_\Gg$ if it is rotated so that its
$\tau\tau$ line consists of short rods.
In other words, tensor of eigenstrain $\bE_\Gd$ must be rotated to a proper angle  so that its tangent
entry becomes compatible with $\bE_\gamma$ \eq{bas3}, $\left[\bE_\Gd\right]_{\tau \tau}=1$, see compatibility
conditions \eq{comp11}.  The rotated matrix is computed from the additional conditions that trace and
determinant are independent of rotation. The rotated
eigenstrain becomes
$${\bE_\Gd}\sub{(rotated)}=
\begin{pmatrix}
\f{2}{\sqrt{3}}\sin(b)+2\cos(b)-1 & \Gr \cr \Gr & 1
\end{pmatrix}
$$
where $\Gr^2=\f{1}{3}(2\cos(b)-1)(2\sin(b)\sqrt{3}-3)$.
Eigenstrain $\bE_{\Gd\Gg}$ of the mixture is still an average of
those in the phases according to \eq{10}. \beqa \mu
{\bE_\Gd}\sub{(rotated)}+ (1-\mu) \bE_\Gg= \nonum
\begin{pmatrix}
\mu\left(\f{2}{\sqrt{3}}\sin(b)+2\cos(b)-1\right)+ (1-\mu) \f{\tan
\Gg}{\sqrt{3}} & \mu \Gr \cr \mu \Gr & 1
\end{pmatrix}
\eeqa{lam-anis} where $\mu$ is the volume fraction, as before.

Eigenvalues $\GL_{\Gd\Gg}$ of a
composite $\bE_{\Gd\Gg}$ are parametrized by $\mu\in [0,1]$. They depend on $\mu$ nonlinearly.
In the eigenvalue plane, they are located at an arch that spans $\GG_\Gd$
and $\GL_\Gg$, not at an interval between them.

\subsubsection{Two-phase twins} Besides these structures, there are
two-phase twins: the laminates of differently oriented phase $\bE_\Gd$
or $\bE_\Gg$. These structures preserve the trace of $\bE_\Gd$ and
$\bE_\Gg$, respectively, and their eigenstrain is as follows
$${\bE_\Gd}\sub{(twin)}= \begin{pmatrix}  \f{2 }{\sqrt{3}}
\sin(b)+2\cos(b)-1   & (1-2\mu) \Gr \cr (1-2\mu)\Gr & 1
\end{pmatrix}
$$
where $ \mu =k/N$, $k\in [-N, N]$. One could check that the eigenvalues of these
twins are never isotropic, Figure \ref{fig:laminates_tw}.

\begin{figure}
\begin{center}
\psfrag{gamma}{$\gamma$}\psfrag{alpha}{$\alpha$}\psfrag{beta}{$\beta$}\psfrag{eta}{$\eta$}\psfrag{delta}{$\delta$}
\includegraphics[width=0.5\textwidth]{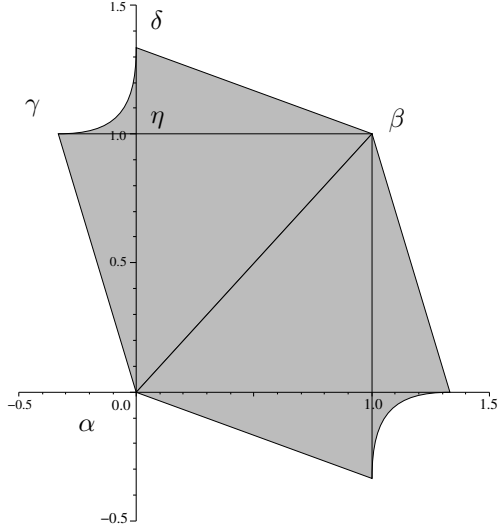}
\caption{Region of eigenvalues of still states}
\label{fig:region_eigenvalues}
\end{center}
\end{figure}

Together, all listed structures form a region $\GO$ in the plane of
eigenvalues of $\bE$. The boundary has three components corresponding
to  structures $ S_{\Ga\Gd}$ (an interval), $ S_{\Gd\Gg}$ (an arch),
and $ S_{\Gg\Gb}$ (an interval) and four vertices, see Figure \ref{fig:region_eigenvalues}.

\subsubsection{Three-phase laminates} There exist three-phase still
structures with eigenstrains that densely cover the domain
$\GO$ bounded by the eigenstrains of two-phase structures.
Curvilinear triangular domain $\Ga, \Gd,\Gg$ is covered with
laminates from three phases, obtained by laminating the isotropic
phase $S_\Ga$ with the laminate $S_{\Gd\Gg}$. The tangents to
laminates correspond to the sides of unit lengths in all three
components. The eigenstrain $\bE_{\Ga \Gd\Gg}$ is
$$\bE_{\Ga \Gd\Gg}=\nu \bE_\Ga+ \mu {\bE_\Gd}\sub{(rotated)}+ (1-\mu-\nu)
\bE_\Gg
$$
where
$$
\mu=\f{k}{N}, ~k=0, \ldots, N, \quad \nu=\f{p}{N}, ~p=0, \ldots,
N-k,
$$
The corresponding eigenvalue pairs cover the part of $\Omega$ that lies outside
the rectangle, three of whose  vertices are $\alpha, \eta, beta$ in Figure \ref{fig:region_eigenvalues}.

\subsection{Hexagon-triangles-strips}
\begin{figure}
\begin{center}
	\subfigure[Step 1]{\includegraphics[height=0.3\textwidth]{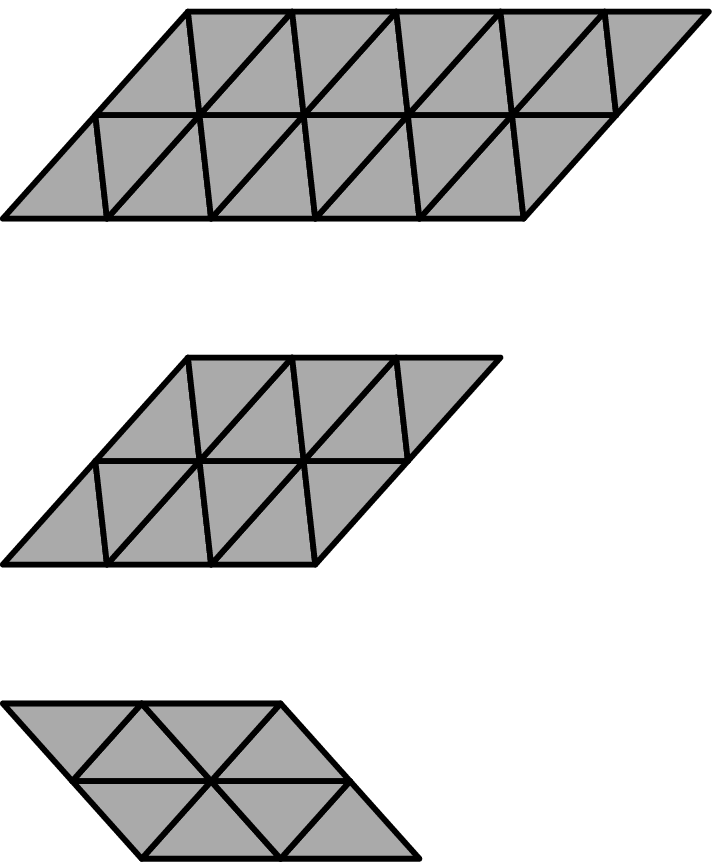}}
	\subfigure[Step 2]{\includegraphics[height=0.3\textwidth]{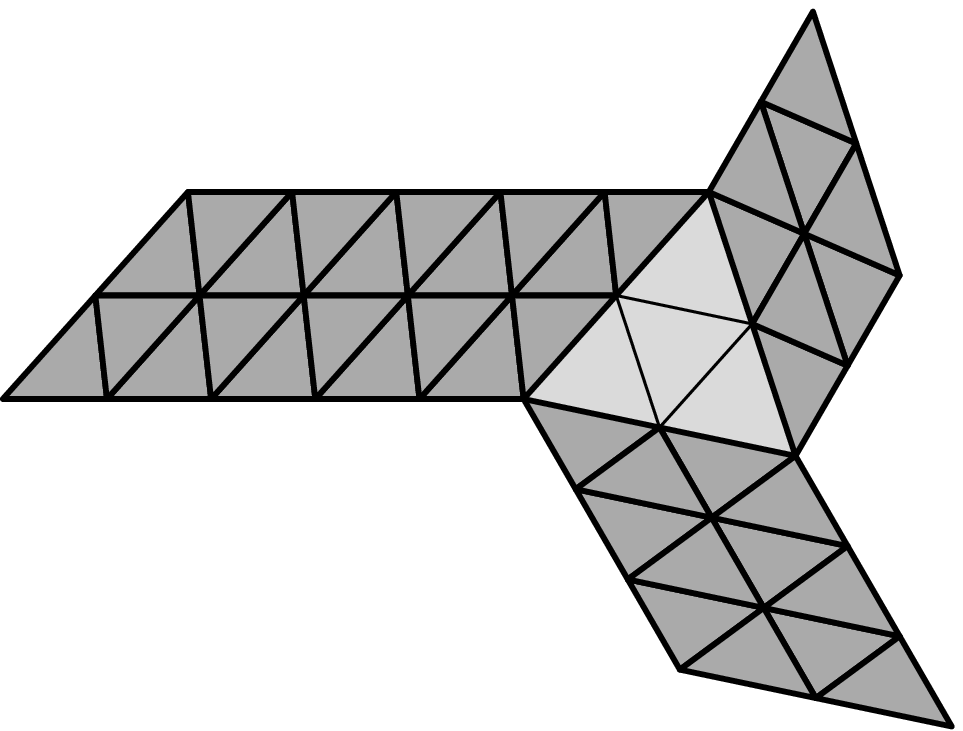}}
	\subfigure[Step 3, the structure]{\includegraphics[width=0.8\textwidth]{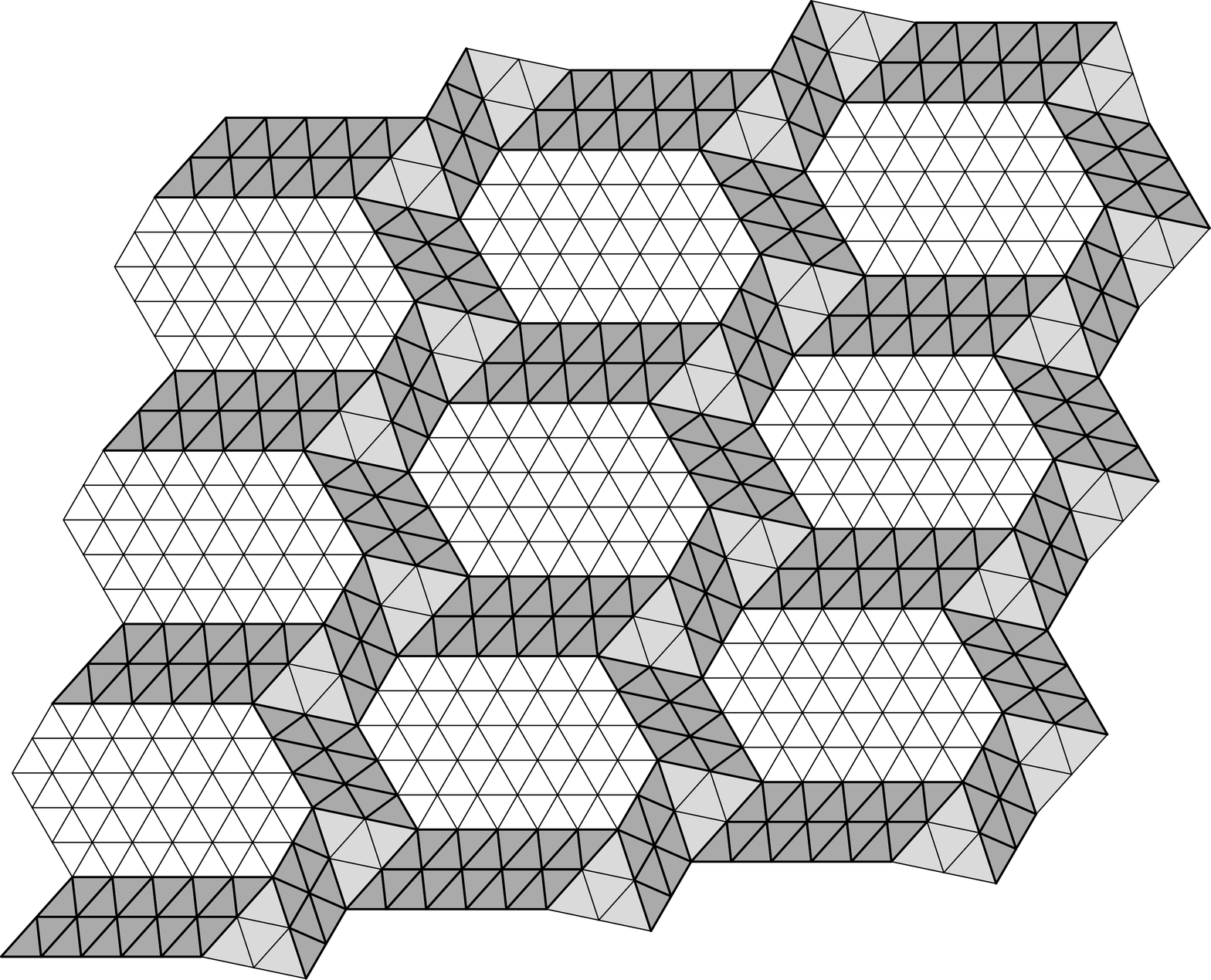}}
\end{center}
\caption{Hexagon-triangle strips and the assembly}
\label{fig:hex_strips}
\end{figure}

 The central triangle $\Ga, \eta, \Gb$ can be covered by
mosaic assembly from elements of $S_\Ga$, $S_\Gb$ and differently
oriented elements $S_\Gd$, $S_\Gd'$  $S_\Gd''$ shown in Figure
\ref{fig:hex_strips}. Notice that  $S_\Ga$ and $S_\Gb$ are incompatible, and
the assembly uses anisotropic elements  $S_\Gd$ between them. The
assembly is as follows.
\begin{enumerate}
\item First, three parallelograms from  $S_\Gd$, $S_\Gd'$ and
$S_\Gd''$ are formed of the sizes $k, n_1$, $k, n_2$, and $k, n_3$,
respectively. The links of the unit length form the base of each
parallelogram and the links of the length $a$ form the sides (Figure \ref{fig:hex_strips}, (a)).
\item
Second, these parallelograms are joined by the edges with the
equilateral triangle with the side $k a$ in between, forming the
structure shaped as Y (Figure \ref{fig:hex_strips}, (b)).
\item
 Third, the obtained figure is copied and
translated. The copies are joined by edges, leaving empty triangles
and hexagons with the parallel sides of $n_1$, $n_2$ and $n_3$
unit-length elements.
\item
Finally, the triangular parts are filled with $S_\Gd$ and the hexagon
parts with $S_\Ga$. The obtained periodic assembly covers the whole
plane (Figure \ref{fig:hex_strips}, (c)).
\end{enumerate}The obtained periodic assembly covers the whole plane.

 The periodicity element consists of

-- three parallelograms from  $S_\Gd$, $S_\Gd'$  $S_\Gd''$
respectively, of $2n_1\, k$, $2n_2\, k$, and $2n_3\, k$ triangles
each,

-- two triangles composed of $S_\Gg$ of $  k(k+1)$ triangles each.

-- one hexagon containing $S_a$ of $ 2 (n_1 n_2+ n_2 n_3+ n_3 n_1)$
triangles.

Thus the element consists of $$N=k(k+1+ n_1+n_2+n_3) +n_1 n_2+ n_2 n_3+ n_3 n_1$$
pairs of triangles.

The eigenstrain $\bE\sub{hex}$ of the assembly is a sum of
eigenstrains of its pieces,
 \beq
\bE\sub{hex}=\f{D}{N}
 \eeq{def}
where
 \beq
 D= k(k+1)\bE_a+n_1\, k \bE_\Gd+ n_2\, k \bE_\Gd'+
n_3\, k \bE_\Gd''+(n_1 n_2+ n_2 n_3+ n_3 n_1)\bE_\Gb
 \eeq{D}
$$
\bE\sub{hex}=\f{k(k+1)\bE_a+n_1\, k \bE_\Gd+ n_2\, k \bE_\Gd'+ n_3\, k
\bE_\Gd''+(n_1 n_2+ n_2 n_3+ n_3 n_1)\bE_\Gb}{k(k+1+ n_1+n_2+n_3) +n_1
n_2+ n_2 n_3+ n_3 n_1}
$$

\subsubsection{Asymptotics}

In particular, $\bE\sub{hex}$ is isotropic if $n_1=n_2=n_3=n$. Then
 \beqa
\bE\sub{hex}&=&\Ge\sub{is} I,\nonum \Ge\sub{is}&=&   \f{ k(k+1)+
\f{3}{2} n\, k \Tr (\bE_\Gd)+3 n^2 a}{k(k+1) + 3 n +3 n^2}
 \eeqa{is}
When $n\to \infty$ and $\f{k}{n}$ varies in $[0, \infty]$, the isotropic deformation
$\Ge\sub{is}$ varies from one to $a$.

The described structures degenerate into two-phase structures
$\bE_{\Gd\Gb}$ and $\bE_{\Ga\Gd} $.
 If $n_1 \gg n_2, n_3, k$, then
$$
\bE\sub{hex}\to \f{  k \bE_\Gd + (n_2+n_3) \bE_\Gb}{k+ n_2+ n_2 +
n_3}=\bE_{\Gd\Gb}
$$
If $n_2\ll n_1, k$ and $n_3\ll n_1, k$, we have
$$
\bE\sub{hex}\to \f{k(k+1)\bE_a+n_1\, k \bE_\Gd}{k(k+1)+ n_1 k}=
\bE_{\Ga\Gd} $$
\subsection{Covering of the whole domain $\GO$}
Together, the hexagonal and three-phase-laminate structures densely fill the whole curved
hexagon of eigenvalues of still states. The two types of covers are
met on the line $\Ga, \Gg$ when $k_h\to 1$ in hexagons and $\mu\to
0$ in three-phase laminates. $ \GL_\Ga \GL_\Gd \GL_\Gg \GL_\Gb$. The
covering is non-unique: For instance, one can replace $S_\Ga$ with $
S_\Gb$ and $S_\Gd$ with $S_\Gg$ everywhere (complementary or dual
replacement) and obtain another  complete coverage. The density of
the coverage is easy to estimate through the explicitly given
parameters $k, p$. The distance from an eigenstrain of a
discrete $N\times N$ periodic structure to the arbitrary point of
the domain is of the order of $N^{-1}$, as follows from the above
explicit formulas.


\section{Small deformations}
\subsection{Linearized elongations}
In the remainder of this paper we consider the case of small deformations which means that rod elongations are linearized near the reference configuration. The linearized relative elongation $\kappa_{ij}$ of the rod $(i, j)$ can be written as
\begin{equation}
\label{kappa}
\kappa_{ij} = \bq_{ij} \cdot \frac{\bu_j - \bu_i}{l},
\end{equation}
where $\bq_{ij}$ is the unit direction from
node $i$ to node $j$, $\bu_i$ and $\bu_j$ are the displacements of
nodes $i$ and $j$, respectively.

Consider a triangular periodic network of $N$ nodes, where $N$ is finite, even though it can be arbitrary large.
We also suppose that the nodes are arranged into a hexagonal shape containing $n$ nodes along a side
(see Figure 5.1 for an example).

\begin{figure}[htp]
\begin{center}
\psfrag{w}{$w$} \psfrag{s}{$s$} \psfrag{kappa}{$\kappa$} \psfrag{0}{$0$}
\includegraphics[width=3in]{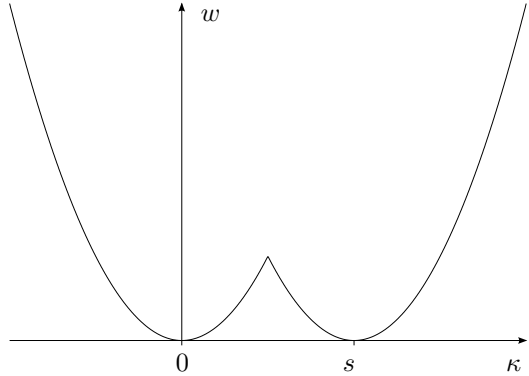}
\caption{Energy of a single rod}
\label{fig:link_energy}
\end{center}
\end{figure}

Factoring out $l^2$ in (\ref{rod-energy}) and using nodal displacements instead of the rod length, we write the energy of a rod $(i,j)$ as
\begin{equation}
\label{spring-energy}
 w(\kappa_{ij}) = \frac{C l^2}{2} \min \{\kappa_{ij}^2, (\kappa_{ij} - s)^2\}, \qquad
 \end{equation}
where, $s>0 $ is a dimensionless parameter characterizing the
critical relative elongation. The graph of the dependence
$w(\kappa_{ij})$ is shown on figure \ref{fig:link_energy} (Compare with (\ref{rod-energy}), where $x$ denotes the actual rod length).

The displacements $\Bu_k$ in the equilibrium state are found by
minimizing the total energy of the network:
$$ W(\bu_1, \ldots \bu_N) = \sum_{(i,j)} w(\kappa_{ij}) = \sum_{(i,j)} w\left(\bq_{ij} \cdot \frac{\bu_j - \bu_i}{l}\right) $$

\subsection{Average strain}
An effective homogeneous deformation state of the network is characterized by an average (linearized) strain tensor $\bE$. In this section we obtain a formula for the average strain in terms of elongations $\kappa_{ij}$.

Denote the number of nodes in the network by $N$.
Given a set of displacements of nodes $\bu_1, \ldots \bu_N$ we define a continuum deformation
$\bu(x)$ that coincides with $\bu_i$ at each $\bx_i$. This can be done by interpolating. In the present case, the lattice
forms a triangulation of the physical domain $\Omega$. Therefore it is easy to construct a piecewise linear interpolant, using finite elements. After this is done we obtain
a continuous function $\bu(\bx), \bx\in \Omega$ such that
\begin{itemize}
\item $\bu(\bx_i) = \bu_i$ for $i = 1, \ldots N$ and $\bx_i$ being the position of node $i$;
\item $\bu(\bx)$ is linear on every elementary triangle of the lattice.
\end{itemize}

To $\bu(\bx)$ we associate strain tensor $\bvep$:
$$ \bvep = \frac{\nabla \bu + (\nabla \bu)^T}{2} $$

Average strain $\bE$ is defined by the formula
\begin{equation} \label{eq:astrain_def} \bE = \frac{1}{|\Omega|} \int_\Omega \bvep(\bx) d\bx, \end{equation}
where $\Omega$ is the domain of $\bu$ and $|\Omega|$ is
$$ |\Omega| = \int_\Omega d\bx $$

There exist formulas for $\bE$ in terms of displacements $\bu_i$ (see \cite{stress_and_strain_bagi}, \cite{t_definitions_sataki}).
However, for our purposes
it is better to work with the relative elongations $\kappa_{ij}$ because we would often prescribe elongations, rather than displacements. To use standard formulas one would have to solve the system (\ref{r-system}) (see the next section) that relates elongations and displacements. We choose a more direct route that involves relating three average elongations along the lattice directions with the three independent components of the average strain.

Rewrite (\ref{eq:astrain_def}) as
$$ \bE = \frac{1}{|\Omega|} \int_\Omega \bvep(\bx) d\bx = \frac{1}{|\mathcal T|} \sum_{\Delta \in \mathcal T} \bvep_\Delta, $$
where $\mathcal T$ is the set of elementary triangles of the network, $|\mathcal T|$ is the number of elementary triangles
in the network, and $\bvep_\Delta$ is the strain tensor on elementary triangle $\Delta$:
$$ \bvep_\Delta = \frac{\nabla \bu_\Delta + (\nabla \bu_\Delta)^T}{2} = \frac{1}{|\Omega_\Delta|} \int_{\Omega_\Delta} \bvep(\bx) d\bx $$

Since the average strain tensor $\bE$ is a $2 \times 2$ symmetric matrix, it has three independent components:
$$
\bE = \left[\begin{array}{cc}
a & b \\
b & c
\end{array}\right]
$$

\begin{figure}[htp]
\begin{center}
	\subfigure[Three lattice directions]{
		\psfrag{q1}{$\bq_1$}\psfrag{q2}{$\bq_2$}\psfrag{q3}{$\bq_3$}\psfrag{u1}{$\bu_1$}\psfrag{u2}{$\bu_2$}
		\includegraphics[width=0.3\textwidth]{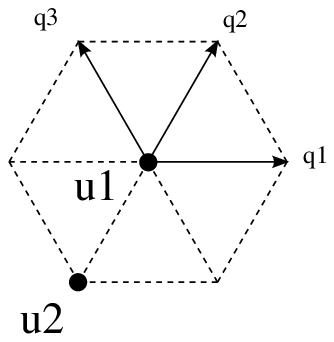}
		\label{fig:three_directions}
	}
	\subfigure[Definition of vector $\bk_\Delta$]{
		\psfrag{q1}{$\bq_1$}\psfrag{q2}{$\bq_2$}\psfrag{q3}{$\bq_3$} \psfrag{i}{$i$}\psfrag{j}{$j$}\psfrag{h}{$h$}
		\includegraphics[width=0.3\textwidth]{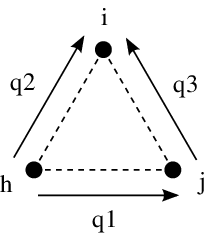}
		\label{fig:k_delta}
	}
\caption{}
\end{center}
\end{figure}

Consider three lattice direction vectors $\bq_1$, $\bq_2$, and $\bq_3$ (see figure \ref{fig:three_directions}):
$$
\bq_1 = \frac{1}{2}\left[\begin{array}{r} 2 \\0 \end{array}\right], \quad
\bq_2 = \frac{1}{2}\left[\begin{array}{r} 1 \\ \sqrt 3 \end{array}\right], \quad
\bq_3 = \frac{1}{2}\left[\begin{array}{r} -1 \\\sqrt 3 \end{array}\right],
$$

We can find the components of $\bE$ using the values $\bq_r \cdot (\bE \bq_r)$, $r=1,2,3$. Hence, consider
\begin{equation}
\label{m1}
\bq_r \cdot (\bE \bq_r) = \frac{1}{|\mathcal T|} \sum_{\Delta \in \mathcal T} \bq_r \cdot (\bvep_\Delta \bq_r)
= \frac{1}{|\mathcal T|} \sum_{\Delta \in \mathcal T} \bq_r \cdot (\nabla \bu_\Delta \bq_r) .
\end{equation}

Using linearity of $\bu$ on triangle $\Delta$ for edge $(i,j)$ of the triangle we find
$$ \bq_{ij} \cdot (\bvep_\Delta \bq_{ij}) = \bq_{ij} \cdot (\nabla \bu_\Delta \bq_{ij}) = \bq_{ij} \cdot \frac{\bu_j - \bu_i}{l} = \kappa_{ij} $$

Given a triangle $\Delta$ with vertices $h,i,j$ (see figure
\ref{fig:k_delta}) define vector $\bk_\Delta \in \mathbb R^3$ as:
$$ \bk_\Delta = \left[\begin{array}{c}
\kappa_{hj} \\
\kappa_{hi} \\
\kappa_{ij}
\end{array}\right] $$

Using this notation we rewrite (\ref{m1}) as
a linear system for finding the components $a,b,c$ of $\bE$ given elongations $\kappa_{ij}$:
\begin{equation} \label{eq:astrain_system}
Q
\left[\begin{array}{c}
a \\
b \\
c
\end{array}\right] = \frac{1}{|\mathcal T|}\sum_{\Delta \in \mathcal T} \bk_\Delta,
\end{equation}
where the matrix $Q$ is given by
\begin{equation}
\label{eq:matrix_q}
Q =
\left[\begin{array}{rrr}
q_{1, 1}^2 ~~~~~& 2q_{1, 1} q_{1, 2}~~~~~ &  q_{1, 2}^2\\
q_{2, 1}^2 ~~~~~ & 2q_{2, 1} q_{2, 2}~~~~~ &  q_{2, 2}^2\\
q_{3, 1}^2~~~~~   & 2q_{3, 1} q_{3, 2}~~~~~ & q_{3, 2}^2
\end{array}\right]=
\left[\begin{array}{rrr}
1 & 0 & 0 \\
\frac{1}{4} & \frac{\sqrt 3}{2} & \frac{3}{4} \\
\frac{1}{4} & -\frac{\sqrt 3}{2} & \frac{3}{4}
\end{array}\right]
\end{equation}
and function $\m:\mathbb R^3 \to \mathbb R^{2\times2}$ by
$$ \m(\bx) = \bx_1 \left[\begin{array}{cc}
1 & 0 \\
0 & 0
\end{array}\right] + \bx_2 \left[\begin{array}{cc}
0 & 1 \\
1 & 0
\end{array}\right] + \bx_3 \left[\begin{array}{cc}
0 & 0 \\
0 & 1
\end{array}\right] $$
Function $\m$ is simply a linear 1-1 mapping from the space of three dimensional
vectors to the space of $2\times 2$ symmetric matrices.

Solving (\ref{eq:astrain_system}) we find average strain $\bE$:
\begin{equation} \label{eq:astrain_expr}
\bE = \frac{1}{|\mathcal T|} \m\left(Q^{-1}\sum_{\Delta \in \mathcal T} \bk_\Delta\right),
\end{equation}
where
$$ Q^{-1} = \left[\begin{array}{rrr}
1 & 0 & 0 \\
0 & \frac{\sqrt 3}{3} & -\frac{\sqrt 3}{3} \\
-\frac{1}{3} & \frac{2}{3} & \frac{2}{3}
\end{array}\right] $$

Further, we will rewrite the sum over triangles in
(\ref{eq:astrain_expr}) as the sum over edges (this is just the change in indexing). Denote all edges of
the array by $\mathcal E$, all boundary edges of the array by
$\mathcal E_B$ and all non-boundary edges of the array by $\mathcal
E_I$ ($\mathcal E_I = \mathcal E \backslash \mathcal E_B$). Then
$$ \sum_{\Delta \in \mathcal T} \bk_\Delta = 2 \sum_{e \in \mathcal E} \bk_e - \sum_{e \in \mathcal E_B} \bk_e, $$
where vector $\bk_e \in \mathbb R^3$ has the components defined by
$$ (\bk_e)_r = \left\{\begin{array}{ll}
\kappa_e, & \mbox{ if edge $e$ is parallel to $\bq_r$} \\
0, & \mbox{ else}
\end{array}\right. , \quad r = 1, 2, 3 $$
Next, equation (\ref{eq:astrain_expr}) can be rewritten as
$$ \bE = \frac{2}{|\mathcal T|} \m\left(Q^{-1}\sum_{e \in \mathcal E} \bk_e\right) - \frac{1}{|\mathcal T|} \m\left(Q^{-1}\sum_{e \in \mathcal E_B} \bk_e\right) .$$

Denote the average elongation by $\bar \bk$:
\begin{equation}\label{eq:average_elong}
\bar \bk = \frac{1}{|\mathcal E|} \sum_{e \in \mathcal E} \bk_e .
\end{equation}
Then we rewrite (\ref{eq:astrain_expr}) and obtain the desired formula for the average strain in terms of average elongations:
\begin{equation}\label{eq:astrain_expr2}
\bE = 2\frac{|\mathcal E|}{|\mathcal T|} \m(Q^{-1}\bar \bk) - \frac{1}{|\mathcal T|} \m\left(Q^{-1}\sum_{e \in \mathcal E_B} \bk_e\right).
\end{equation}
This formula is used extensively in the remainder of the paper. The first term in the right hand side contains contributions from all interior edges of the lattice, while the second term contains the contributions from the boundary edges only. As the number of nodes increases, the second term becomes a small perturbation of the first. This becomes clear from the following estimate
\begin{equation}
\label{eq:average_strain_tail}
\left\| m\left(\frac{1}{|\mathcal T|} Q^{-1} \sum_{e \in \mathcal E_B} \bk_e \right)\right\|_F \leq \sqrt{2} \frac{|\mathcal E_B|}{|\mathcal T|} ||Q^{-1}||_2  \max_{e \in \mathcal E_B} ||\bk_e||_2.
\end{equation}
Here $\parallel \cdot\parallel_F$ denotes the Frobenius norm of a matrix, and $\parallel\cdot\parallel_2$ is the Euclidean norm. Note that the ratio
$$
\frac{|\mathcal E_B|}{|\mathcal T|}
$$
of the number of the boundary edges to the number of all elementary triangles goes to zero as the number of nodes approaches infinity.

\section{Compatibility conditions}
\label{section:hex_equations}
In the next two sections, we study the linearized version of the problem of finding still states. The linearized
formulation is as follows. Equations (\ref{kappa}) relating node displacements with link elongations can be concisely
written as
\begin{equation}
\label{r-system}
R\bU = \bkap ,
\end{equation}
where $\bU$ is a tuple of displacements $\bu_i$, say $(\bu_1,
\ldots \bu_N)$, and $\bkap$ is a tuple of $\kappa_{ij}$. The number
of unknowns in (\ref{r-system}) is $2N$ (since
each $\bu_i$ has two independent components). The number of equations
is larger,  approximately $3N$ (see Lemma 5.1 below for the exact count). Therefore, we have an overdetermined
system (the number of unknowns is less than the number of
equations). As a consequence, (\ref{r-system}) may be
unsolvable for some $\bkap$. The values of $\bkap$ for which
(\ref{r-system}) is solvable are called {\it admissible}.

\subsection{Compatibility conditions characterize range of $R$}

 In this section we show that
compatibility conditions (\ref{comp6}) ar necessary and sufficient for solvability of
(\ref{r-system}). Necessity is clear from the derivation of (\ref{comp6}), so we focus on sufficiency.

\begin{figure}[htp]
\begin{center}
    \includegraphics[height=2in]{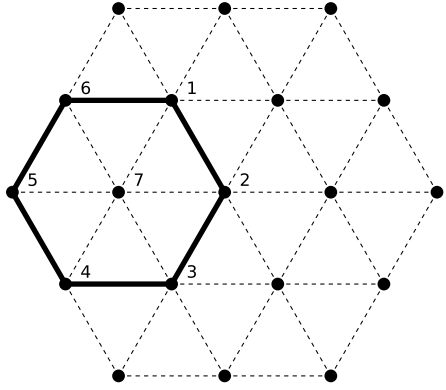}
\end{center}
\caption{Example of the network with $N=19$, $n=3$. A hexagon that forms a hexagonal equations shown in bold}
\label{fig:sample_array}
\end{figure}

Suppose that we have the hexagonal array with a side of $n$ points (see figure \ref{fig:sample_array}).

For every hexagon consisting of 6 vertices we formulate the condition (\ref{comp6}): the sum of elongations
of the outer edges equals to the sum of elongations of the inner edges.
For example, for figure \ref{fig:sample_array}
the equation will be
$$ \kappa_{12} + \kappa_{23} + ... + \kappa_{56} + \kappa_{61} = \kappa_{71}+\kappa_{72}+...+\kappa_{75}+\kappa_{76} $$
The collection of all such equations will be called {\em hexagonal equations}.

Next, we need a preliminary result giving the exact count of the number of nodes, edges and hexagonal equations. The following lemma is used later to carry out proofs by induction.

\begin{lemma} The total number of nodes $N$, the number of edges $E$,
and the number of hexagonal equations $M$ depend on $n$ as follows.
$$ N(n) = 3 n^2 - 3 n + 1 , \qquad E(n) = 9 n^2 - 15 n + 6 , \qquad M(n) = 3 n^2 - 9 n + 7 . $$
\end{lemma}
\begin{proof}
We will prove the lemma using mathematical induction.

For $n = 2$ we have a hexagon that has 7 vertices, 12 edges, and 1 hexagonal equation. Indeed,
$$ N(2) = 3 \cdot 2^2 - 3 \cdot 2 + 1 = 7, \quad E(2) = 9 \cdot 2^2 - 15 \cdot 2 + 6 = 12, \quad M(2) = 3 \cdot 2^2 - 9 \cdot 2 + 7 = 1 $$
Thus, the formulas are true for $n=2$.

Suppose that given formulas are true for $n=k$. Let's prove that they are true for $n = k+1$.

When we increase $n=k$ by 1 we ``wrap'' the array by a layer of points. Therefore, we add
$6k$ vertices, we allow $6(k-1)$ new hexagonal equations, and add $6(3k-1)$ edges.
Thus, the number of vertices $N^*$ for the size $n = k+1$ is:
\begin{eqnarray*}
N^* &=& N(k) + 6k = 3 k^2 - 3 k + 1 + 6k = 3 (k+1-1)^2 + 3 k + 1 \\
&=& 3 (k+1)^2 - 6(k+1) + 3 + 3k + 1 = 3 (k+1)^2 - 3 (k+1) + 1,
\end{eqnarray*}
the number of equations $M^*$ is
\begin{eqnarray*}
M^* &=& M(k) + 6(k-1) = 3 k^2 - 9 k + 7 + 6(k-1) = 3 (k+1-1)^2 - 3 k + 1 \\
&=& 3 (k+1)^2 - 6(k+1) + 3 - 3k + 1 = 3 (k+1)^2 - 9 (k+1) + 7,
\end{eqnarray*}
and the number of edges $E^*$ is
\begin{eqnarray*}
E^* &=& E(k) + 6(3k-1) = 9 k^2 - 15 k + 6 + 18k - 6 = 9 (k+1-1)^2 + 3 k \\
&=& 9 (k+1)^2 - 18(k+1) + 9 + 3k = 9 (k+1)^2 - 15 (k+1) + 6,
\end{eqnarray*}
Since $N^* = N(k+1)$, $M^* = M(k+1)$, and $E^* = E(k+1)$, the formulas are true.
\end{proof}

Observe that
$$2 N - 3 = E - M . $$

We know that for the mapping $R: \mathbb R^{2N} \to \mathbb R^E$ we have $\dim (\ker R) = 3$. This follows from the well known results on graph rigidity, in particular theorems on the first-order rigidity of triangulations
(see e.g. \cite{ABP} and references therein).
Thus,
$$ \dim(\mathrm{im}\ R) = 2N-3 $$

The hexagonal equations can be concisely written in the form
\begin{equation}
\label{zkappa} Z \bkap = 0 ,
\end{equation}
where $Z$ is a matrix, $Z: \mathbb R^E \to \mathbb R^M$; $i$-th row of this matrix corresponds to $i$-th equation.
\begin{lemma}
Matrix $Z$ has full rank $M$.
\end{lemma}
\begin{proof}
We will restate the lemma in the form: all hexagonal equations are linearly independent.

We will prove the lemma in its new form using mathematical induction on $n$. For $n=2$ we have $M=1$. One
equation is linearly independent.

Suppose, the lemma is true for $n=k$. Let's prove it for $n=k+1$.

Increasing $n=k$ by 1 we wrap a hexagon of side size $k$ with a layer of points. Doing so we add some number
of hexagonal equations to those that we had in the original hexagon. The hexagonal equations of the original
hexagon are linearly independent. Each new hexagonal equation contains an edge that is present in no
other equation. Thus, all equations are linearly independent.
\end{proof}

Since matrix $Z$ has rank $M$, we find $\dim(\ker Z) = E-M$. Thus,
\begin{equation}
\label{eq_hex_dim}
\dim(\im R) = \dim(\ker Z)
\end{equation}

Necessity of hexagonal equations implies that each admissible $\bkap$ satisfies them. This fact can be stated as
follows:
for any $\bkap \in \im R$ we have $Z \bkap = 0$. This in turn implies
\begin{equation}
\label{eq_hex_subset}
\mathrm{im}\ R \subset \ker Z .
\end{equation}

\begin{theorem}
For any vector $\bkap \in \mathbb R^E$ satisfying $Z \bkap = 0$ there is a unique (up to translation and rotation)
vector $\bu \in \mathbb R^{2N}$ satisfying $R \bu = \bkap$.
\end{theorem}
\begin{proof}
The conditions (\ref{eq_hex_dim}) and (\ref{eq_hex_subset}) imply
$$\im R = \ker Z . $$
Thus, for any $\bkap \in \mathbb R^E$ we can find a vector $\bu \in \mathbb R^{2N}$.
If we neglect rigid rotations and translations of the whole lattice (they form the null space of $R$ by the well known results on rigidity of triangulations, (\cite{ABP} and references therein),  then this vector is unique.
\end{proof}

\section{Still states and small deformation eigenstrains}
In the small deformation case,
a still state is a collection of nodal displacements $\bu_i$)
 satisfying equations (\ref{r-system})
with the right hand side $\bkap$ of special form. The components $\kappa_{ij}$ can take only two values:
$0$ and $s/2$, where $s/2$ is the critical elongation from (\ref{spring-energy}).
If $\kappa_{ij}=0$ we call the corresponding edge $(i, j)$ {\it
short}, otherwise the edge is called {\it long}. To
construct a still state, one could choose
a length for each edge (long or short). The resulting configuration
is accepted if the resulting triangles form a
tessellation. Otherwise the configuration is rejected. In the case of small deformations, an admissible
elongation vector $\bkap$ should lie in the range of the matrix $R$ from (\ref{r-system}). Equivalently, such
$\bkap$ must be a solution of the hexagonal equations (\ref{zkappa}).

Denote the set  of all still states by $\mathcal U$. To characterize  $\mathcal U$,
one needs to solve the following problem.
\begin{quote}
Given $\bkap$ with $\bkap_{ij} \in \{0,s\}$, find $\bU$ realizing that
$\bkap$, that is, find $\bU$such that $R\bU=\bkap$.
\end{quote}
This is possible only if $\bkap$ satisfies hexagonal equations from
Section \ref{section:hex_equations}. Complete
characterization of ${\mathcal U}$ seems very difficult, and is not
addressed in this paper. The reason for the difficulty is the lack
of a convenient structure. It is easily checked that ${\mathcal U}$ is not a linear space, or even a convex set.
Therefore, a description of this set cannot be obtained by linear algebra methods.

 \subsection{An algorithm of adding still states}
In this section we characterize the set of strains that can
be well approximated by the average strains of still states. An average
strain of a still state can be conveniently described by a triple of concentrations of
long edges parallel to the lattice directions. In this section we propose
an explicit construction that furnishes a large number of still
states. Concentrations of these states densely fill a certain unit
cube in the concentration space. Our construction is based on two
observations. First, one needs a binary operation that produces new
still states by combining two already known still states. Second,
one needs a simple "building block", that is, a still state that
could be easily combined with its slightly modified (e.g. translated and rotated)
replicas. In view of the linear structure of $\operatorname{im} R$, the convenient operation
is summation. However, summation of two still states does not always produce a still state.
An additional condition that ensures that the sum of still states remains a still state is
non-overlapping of the long edges (see Lemma \ref{sum_of_ss} below):
if a particular edge is long in the first state, then it should be short in the second state, and visa versa.

From this we deduce that
the building block should have low concentrations of
long edges, and the placement of these edges should be localized as
highly  as possible. If both of these requirements are satisfied,
one can add together shifted copies of the building block to
generate new still states with higher concentrations.

\begin{lemma} \label{sum_of_ss}
The sum $\bkap_1 + \bkap_2$ of two still states $\bkap_1$ and $\bkap_2$
is a still state provided $\bkap_1 \cdot \bkap_2 = 0$.
\end{lemma}
\begin{proof}
Since $\bkap_1$ and $\bkap_2$ are still states, their components satisfy
$$ (\bkap_1)_{ij} \in \{0,s\}, \quad (\bkap_2)_{ij} \in \{0,s\} $$
for all connected nodes $i$ and $j$. Moreover, the condition of orthogonality $\bkap_1 \cdot \bkap_2 = 0$ says that
for each certain pair $(i,j)$ the values $(\bkap_1)_{ij}$ and $(\bkap_2)_{ij}$ cannot equal $s$ simultaneously. Thus,
$$ (\bkap_1)_{ij} + (\bkap_2)_{ij} \in \{0, s\} $$

Also, since $\bkap_1$ and $\bkap_2$ are still states, we have
$$ Z \bkap_1 = 0, \quad Z \bkap_2 = 0, $$ where $Z$ is the matrix of the hexagonal system (\ref{zkappa}).
By linearity $Z$,
$$ Z(\bkap_1 + \bkap_2) = Z \bkap_1 + Z \bkap_2 = 0 $$

Since the necessary conditions hold, the sum $\bkap_1 + \bkap_2$ is a still state.

\end{proof}

Notice that Lemma \ref{sum_of_ss} states that  sum
$$ \bkap_1 + \bkap_2 + \ldots + \bkap_m $$
is a still state, provided $\bkap_i$, $i=1, \ldots, m$ have non-overlapping long edges.

For a still state in (\ref{eq:average_strain_tail}) we have $||\bk_e||_2 \leq s$. Thus
\begin{equation}\label{eq:astrain_tail2}
\left\|\m\left( \frac{1}{|\mathcal T|} Q^{-1} \sum_{e \in \mathcal E_B} \bk_e \right)\right\|_F \leq \sqrt{2}\frac{|\mathcal E_B|}{|\mathcal T|} ||Q^{-1}||_2 s
\end{equation}

\begin{theorem} \label{main_still_theorem}
For any $\alpha_i \in [0,1]$, $i=1,2,3$, and any $n\geq 3$ there exists
a still state $\bkap^*$ such that its concentrations
$\alpha_i^*$, $i=1, 2, 3$, satisfy
$$ |\alpha_i - \alpha_i^*| < \frac{1}{n} $$
\end{theorem}
\begin{proof}
To prove the theorem we will use special still states called {\it stripes}. These
states have $\bkap_{ij} = 0$ everywhere except for one stripe that has non-zero elements in one direction
(figure \ref{fig:sample_still_s}). We can see that these stripes are still states
because their components are either 0 or $s$ and they satisfy the hexagonal equations:
each hexagon has either all zero elongations or one non-zero inner elongation and one non-zero outer
elongation.
\begin{figure}[ht]
\begin{center}
    \subfigure[A horizontal stripe, group 1]{ \label{fig:ss_1} \includegraphics[width=0.47\textwidth]{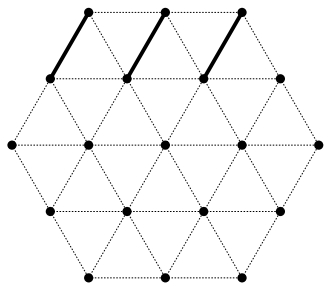} }
    \subfigure[Rotated stripe, group 2]{ \label{fig:ss_2} \includegraphics[width=0.47\textwidth]{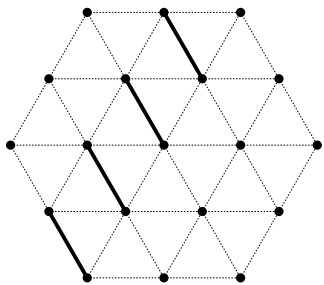} }
    \subfigure[Rotated stripe, group 3]{ \label{fig:ss_3} \includegraphics[width=0.47\textwidth]{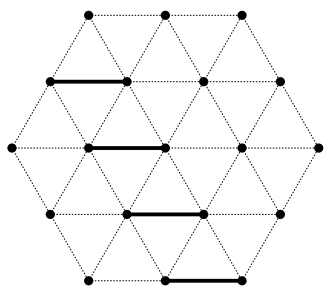} }
    \subfigure[Compound still state]{ \label{fig:cs} \includegraphics[width=0.47\textwidth]{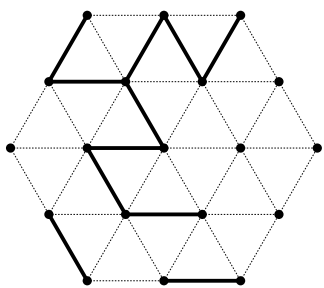} }
\end{center}
\caption{
    \ref{fig:ss_1}, \ref{fig:ss_2}, and \ref{fig:ss_2} are ``stripe'' still states. \ref{fig:cs}
    is a still state composed of \ref{fig:ss_1}, \ref{fig:ss_2}, and \ref{fig:ss_3}.
    Solid line corresponds to $\kappa = s$, dotted line is 0.
}
\label{fig:sample_still_s}
\end{figure}

Observe, that for two different stripes $\bkap_1$ and $\bkap_2$ we have $\bkap_1 \cdot \bkap_2 = 0$. In other words,
the long edges of $\bkap_1$ do not overlap the long edges of $\bkap_2$, and thus $\bkap_1+\bkap_2$ is a still state (figure \ref{fig:cs}).

We have 3 different groups of stripes, $\mathcal G_1$, $\mathcal G_2$, and $\mathcal G_3$, one for each direction:
one group of horizontal stripes and two groups of rotated stripes. The number of stripes in each group is $2(n-1)$.

Next, define the vectors $\bd_r$, $r \in \{1,2,3\}$ as follows. The dimension of $\bd_r$ is $E$ (the number of edges), and
the components are equal to one if the corresponding edge is parallel to $\bq_r$ (a lattice direction vector), and equal to zero otherwise.
It is easy to see that if $\bkap \in \mathcal G_r$ for some $r \in \{1,2,3\}$ then $\bkap \cdot \bd_r < 2 n$,
$r = 1, 2, 3$, since each ``stripe'' has no more than $2n$ long edges.
Also, since each stripe has long edges only in $r$-th direction, we get
$$ \bkap \cdot \bd_r = \frac{E}{3} \alpha_r, \quad \bkap \cdot \bd_t = 0 \mbox{ for } t \neq r. $$
This implies
$$ \alpha_r < \frac{6 n}{E} = \frac{6 n}{9 n^2 - 15 n + 6} < \frac{1}{n} \mbox{ for } n \geq 3. $$

First, we explain the main idea of the proof. Consider $i$-th group (stripes of one certain direction). If we add all stripes of this group
then we'll get $\alpha_i = 1$. For the still state $\bkap = 0$ we have $\alpha_i = 0$. Therefore, adding stripes one by one we
increase $\alpha_i$ from 0 to 1 with steps less than $n^{-1}$. Hence, for any $x \in [0,1]$ there exists a step when $|\alpha_i - x| < n^{-1}$.
Finally, since different directions are independent from each other, we can run this algorithm for $i=1, 2, 3$ and at the end
add all still states together to get what we need. Next, we present the technical details.

For any subset $\mathcal S$ of set $\mathcal G_i$ we will define functions $\hat \bkap(\mathcal S)$ and $\hat \alpha_j(\mathcal S)$ by
$$ \hat \bkap(\mathcal S) = \sum_{\bkap \in \mathcal S} \bkap \quad \mbox{ and }
\quad \hat \alpha_j(\mathcal S) = \sum_{\bkap \in \mathcal S} \frac{3}{E} \bkap \cdot \bd_j = \frac{3}{E} \hat \bkap(\mathcal S) \cdot \bd_j $$

Since $\hat \bkap(\mathcal G_i) = \bd_i$, we have $\hat \alpha_i(\mathcal G_i) = 1$ and
$\hat \alpha_j(\mathcal G_i) = 0$ for $j \neq i$. Also, we define $\hat \bkap(\emptyset) = 0$ and $\hat \alpha_j(\emptyset) = 0$, $j=1,2,3$.

From the corollary of lemma \ref{sum_of_ss} it follows that for any $\mathcal S \subset \mathcal G_i$ the value of function
$\hat \bkap(\mathcal S)$ is a still state with concentrations $\hat \alpha_j(\mathcal S)$.

Suppose we are given values $\alpha_i \in [0,1]$, $i=1,2,3$. Let's define sets $\mathcal H_i$, $i=1,2,3$, as
$$ \mathcal H_i = \{\mathcal S \subset \mathcal G_i : \hat \alpha_i(\mathcal S) \leq \alpha_i\} $$
Having defined $\mathcal H_i$ we define values $\mathcal S^*_i$, $i=1,2,3$, by
$$ \mathcal S^*_i = \begin{array}[t]{c}\mathrm{argmax} \\ ^{\mathcal S \in \mathcal H_i} \end{array} \hat \alpha_i(\mathcal S) $$
Since the sets $\mathcal G_i$ are finite, the sets $\mathcal H_i$ are finite and the maximum exists. Therefore, the definition of $\mathcal S_i^*$
is consistent.

Now if for some $i \in \{1,2,3\}$ we have $\mathcal G_i \backslash \mathcal S^*_i = \emptyset$ then
$\mathcal S^*_i = \mathcal G_i$ and $\hat \alpha(\mathcal S_i^*) = 1$.
From the definition of $\mathcal S_i^*$ we have
$$ \hat \alpha_i(\mathcal S^*_i) \leq \alpha_i $$
Since $\alpha_i \leq 1$, we get $\hat \alpha_i(\mathcal S^*_i)=\alpha_i$.

If for some $i \in \{1,2,3\}$ we have $\mathcal G_i \backslash \mathcal S^*_i \neq \emptyset$
then there exists $\bk \in \mathcal G_i \backslash \mathcal S^*_i$.

From the definition of function $\hat \alpha$ we get
$$ \hat \alpha(\mathcal S_i^* \cup \{\bk\}) = \hat \alpha(\mathcal S_i^*) + \hat \alpha (\{\bk\}) > \hat \alpha(\mathcal S_i^*) $$
And, since $\mathcal S_i^*$ has maximal value in $\mathcal H_i$, we have $\mathcal S_i^* \cup \{\bk\} \not \in \mathcal H_i$.
Hence,
$$ \alpha_i < \hat \alpha_i(\mathcal S^*_i \cup \{\bk\}) $$
Now using
$$ \hat \alpha_i(\mathcal S_i^*) \leq \alpha_i < \hat \alpha_i(\mathcal S^*_i \cup \{\bk\}) $$
and
$$ \hat \alpha_i(\mathcal S^*_i \cup \{\bk\}) - \hat \alpha_i(\mathcal S^*_i) = \hat \alpha_i(\{\bk\}) < \frac{1}{n} $$
we get
$$ 0 \leq \alpha^*_i - \hat \alpha_i(\mathcal S^*_i) < \hat \alpha_i(\mathcal S^*_i \cup \{\bk\}) - \hat \alpha_i(\mathcal S^*_i) < \frac{1}{n} $$

Finally, for the still state $\bkap^*$ defined by
$$ \bkap^* = \hat \bkap(\mathcal S^*_1 \cup \mathcal S^*_2 \cup \mathcal S^*_3) $$
we have
$$ |\alpha^*_i - \alpha_i| < \frac{1}{n} $$
\end{proof}

\begin{figure}[htp]
\begin{center}
    \subfigure[The cube ${[0,1]}^3$]{\includegraphics[width=0.3\textwidth]{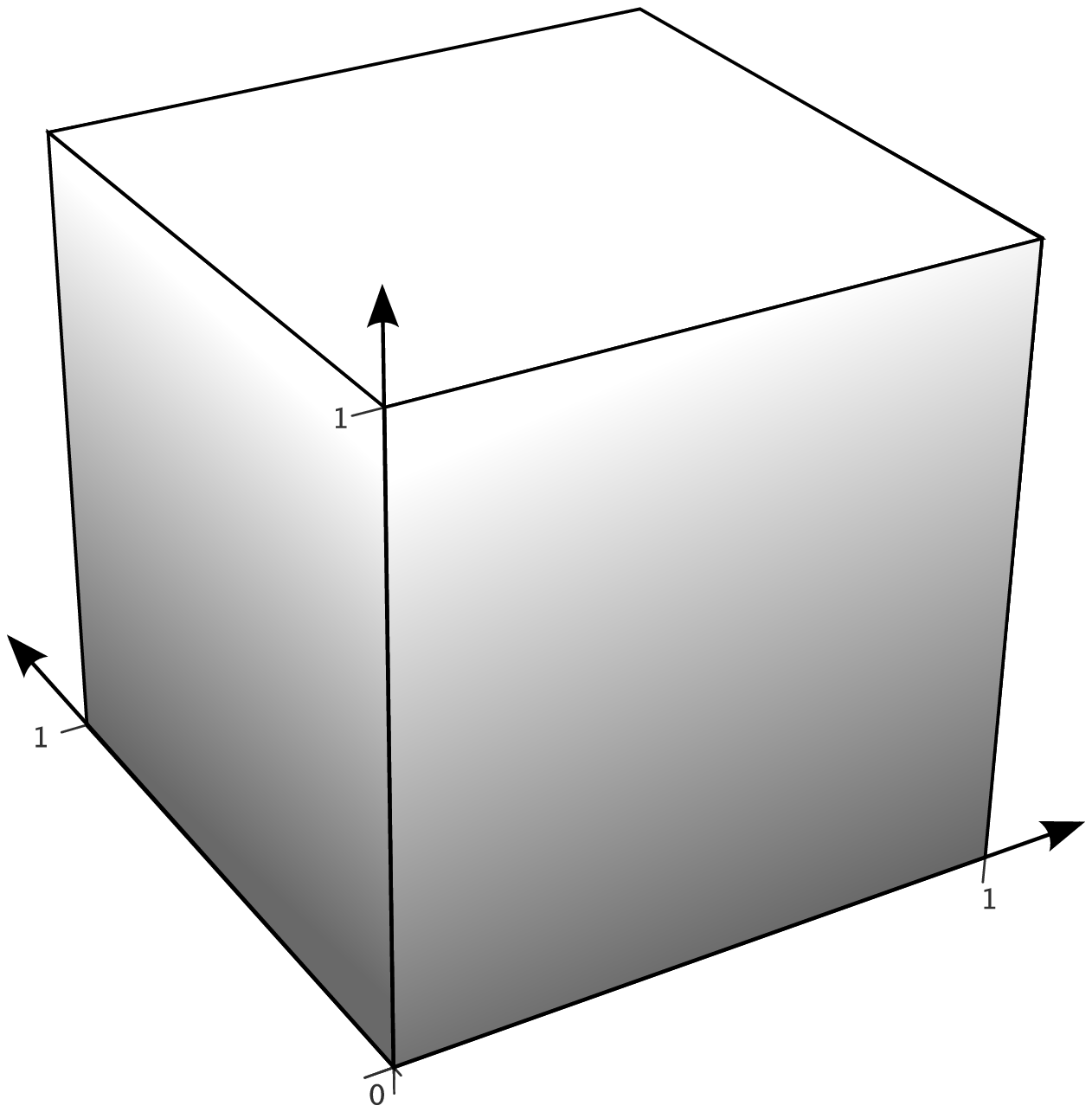} }
    \subfigure[$Q^{-1}{[0,1]}^3$]{\includegraphics[width=0.3\textwidth]{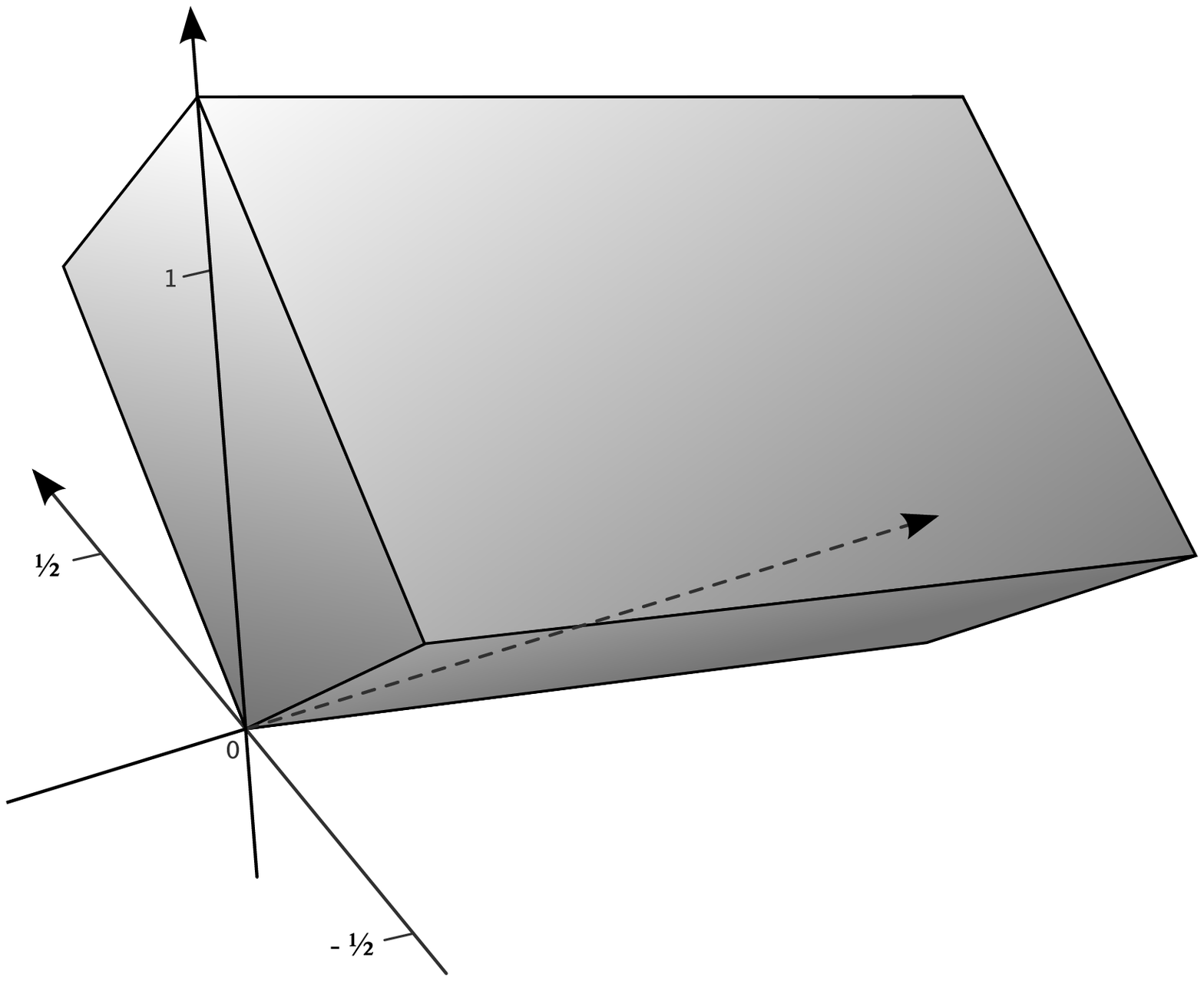} }
    \subfigure[Eigenvalues of $\bar E$]{\psfrag{L_1}{$\lambda_1$}\psfrag{L_2}{$\lambda_2$}\includegraphics[width=0.3\textwidth]{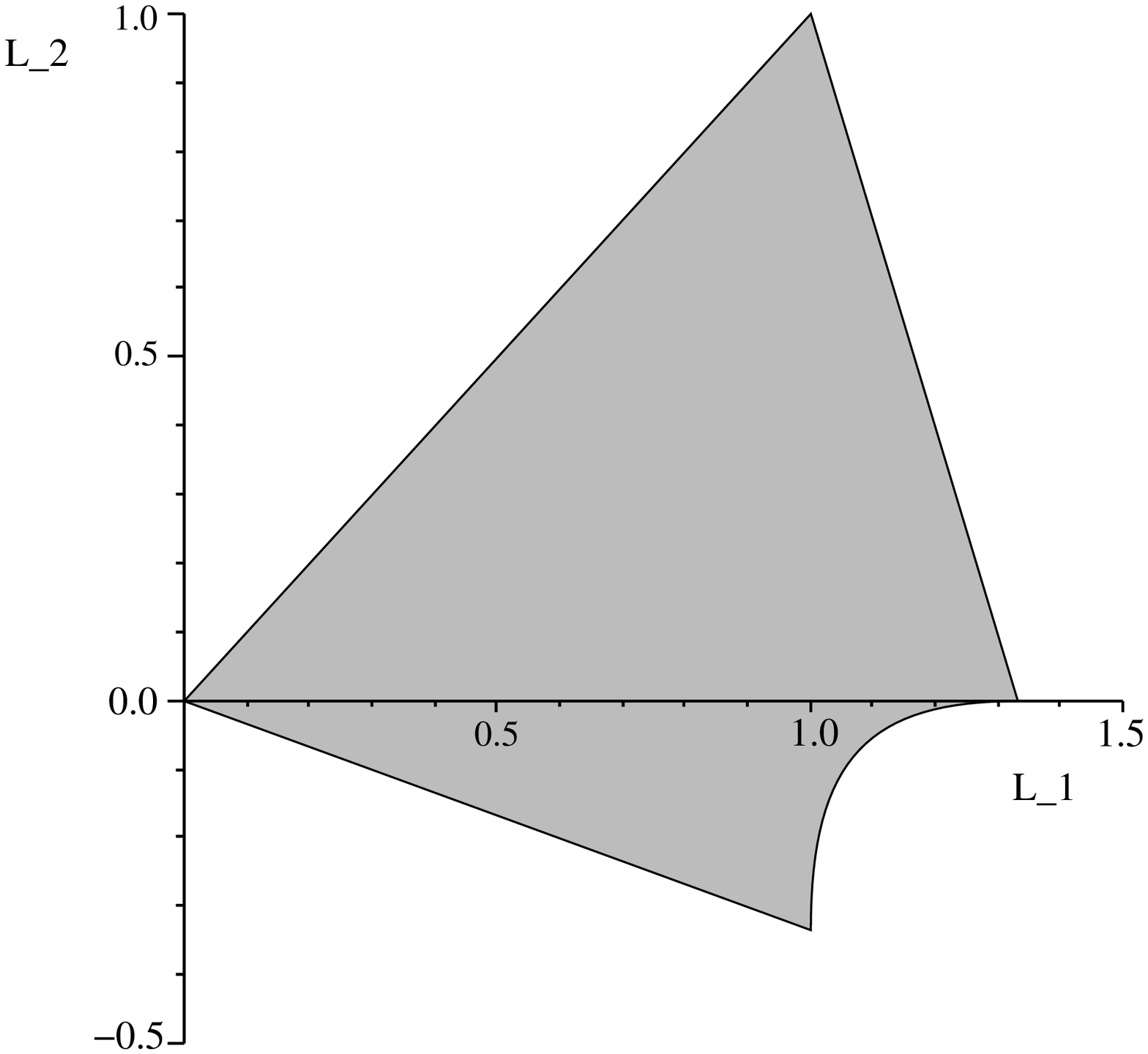} }
\end{center}
\caption{ The cube and its image under $Q^{-1}$ }
\label{fig:cube_image}
\end{figure}

Let us define a set $\mathcal D$:
\begin{equation}
\label{DD}
 \mathcal D = \{s \m(Q^{-1} \bx): \bx \in \mathbb R^3, 0 \leq x_r \leq 1, r = 1, 2, 3 \},
 \end{equation}
 where $Q^{-1}$ is the inverse of the matrix given by Eq. (\ref{eq:matrix_q}).
Set $\mathcal D$ is a convex polygon in the space of $2\times 2$ symmetric matrices, since it is an image
of a cube under a non-singular linear transformation. The components of the elements of $\mathcal D$ form
a set shown on figure \ref{fig:cube_image}.

\begin{theorem}
For any $\bE \in \mathcal D$ there exists a still state of the hexagonal array with a side of $n \geq 3$ points
such that average strain $\bE^*$ of this still state satisfies
$$ ||\bE^* - \bE||_F \leq s ||Q^{-1}||_2 \frac{8}{n-1} $$
\end{theorem}
\begin{proof}
Since $\bE \in \mathcal D$, there exists $\bal \in \mathbb R^3$, $0 \leq \alpha_r \leq 1$, such that
$$ \bE = s \m(Q^{-1} \bal) $$
Using theorem \ref{main_still_theorem} we can claim that for $n \geq 3$ there is a still state with the concentration values $\bal^*$ such that
$$ ||\bal^* - \bal||_2 \leq \frac{\sqrt 3}{n} $$
Hence, using (\ref{eq:astrain_expr2}) we find
\begin{eqnarray*}
||\bE^* - \bE||_F &=& \left\|2 \frac{|\mathcal E|}{|\mathcal T|} \m(Q^{-1} \bar \bk^*) - \frac{1}{|\mathcal T|} \m(Q^{-1} \sum_{e \in \mathcal E_B} \bk_e^*) - \bE\right\|_F \\
&=& \left\|\frac{2s}{3} \frac{|\mathcal E|}{|\mathcal T|} \m(Q^{-1} \bal^*) - \frac{1}{|\mathcal T|} \m(Q^{-1} \sum_{e \in \mathcal E_B} \bk_e^*) - s \m (Q^{-1}\bal)\right\|_F \\
&\leq& \sqrt{2} ||Q^{-1}||_2 \left\|\frac{2s}{3} \frac{|\mathcal E|}{|\mathcal T|} \bal^* - s \bal\right\|_2 + \left\|\m\left(\frac{1}{|\mathcal T|} Q^{-1}\sum_{e \in \mathcal E_B} \bk_e^* \right)\right\|_F
\end{eqnarray*}
Next,
\begin{eqnarray*}
\left\|\frac{2s}{3} \frac{|\mathcal E|}{|\mathcal T|} \bal^* - s\bal\right\|_2 &\leq&
s||\bal^* - \bal||_2 + s\left|1-\frac{2}{3} \frac{|\mathcal E|}{|\mathcal T|}\right| ||\bal^*||_2 \\
&\leq& \frac{\sqrt 3}{n} s + s\left|1-\frac{2}{3} \frac{|\mathcal E|}{|\mathcal T|}\right| \sqrt{3}
\end{eqnarray*}

Further, using
$$ |\mathcal T| = 6 (n-1)^2, \quad |\mathcal E| = 9n^2 - 15n+6, \quad |\mathcal E_B| = 6(n-1) $$
and estimation (\ref{eq:astrain_tail2}) we find
\begin{eqnarray*}
||\bE^* - \bE||_F &\leq& \sqrt{2} ||Q^{-1}||_2 \left(\frac{\sqrt 3}{n} s + s\left|1-\frac{2}{3} \frac{|\mathcal E|}{|\mathcal T|}\right| \sqrt{3}\right)
+\sqrt{2}\frac{|\mathcal E_B|}{|\mathcal T|} ||Q^{-1}||_2 s \\
&=& \sqrt{2} s ||Q^{-1}||_2 \left(\frac{\sqrt 3}{n} + \frac{\sqrt{3}}{3 (n-1)} + \frac{1}{n-1}\right) \\
&\leq& s ||Q^{-1}||_2 \frac{8}{n-1}
\end{eqnarray*}
\end{proof}

Let us now discuss the significance of the Theorem 6.3 for linking the microscopic (lattice-level) and macroscopic (continuum) models.  First, it is of interest to understand how geometry of ${\mathcal D}$ depends on the microstructure. We observe that ${\mathcal D}$ is the
image of a unit cube under a linear mapping defined in (\ref{DD}). The unit cube and the mapping $m$ are both independent of the lattice geometry and particulars of the microscopic force definition. The two microstructure-dependent
quantities in (\ref{DD}) are the relative (non-dimensional) critical elongation $s$ and the matrix $Q^{-1}$.
The actual critical elongation is $sl$, where $l$ is the equilibrium edge length.  Passing to the $\Gamma$-limit requires proper scaling of $l$ with $N$, a typical scaling being $l\sim N^{-1/2}$. In the most natural setting, $sl$ should scale as $l$. If that is the case, then
 $s=O(1)$ as $N\to\infty$. The set ${\mathcal D}$ in that case is independent of $N$, and $s$ is a non-dimensional constant of the problem that determines susceptibility of the material to phase transition. The diameter of ${\mathcal D}$  increases linearly with $s$.
 The matrix $Q^{-1}$ determines orientation of ${\mathcal D}$ in the strain space and provides the explicit dependence of ${\mathcal D}$ on the lattice geometry. Indeed, by definition (\ref{eq:matrix_q}),  $Q$ depends only on the products of components of the lattice direction vectors $\bq_r, r=1, 2, 3$.

 If $s=O(1)$, then ${\mathcal D}$ is the "flat bottom" of the macroscopic energy density because, by Theorem 6.3, any strain in this set can be approximated by a still state eigenstrain and the error
of this approximation vanishes in the limit $N\to\infty$. Since the energy of every still state is zero, and  ${\mathcal D}$ is fixed as $N\to\infty$, the conclusion follows.

\section{Effective energy of the network}
In the previous section, we described the set ${\mathcal D}$ of all  strains that
corresponds to zero effective energy. Here, we suggest the effective energy equation that corresponds
to strains outside this set.

The average elastic energy $W$ of the network without  the transition is
described by Cauchy formula \cite{theory_elasticity_love}
$$
W(\bvep)=\f{2C}{3} \left[\Tr (\bvep^2) - \rec{4} (\Tr \bvep)^2 \right],
$$
where $\bvep$ is the average strain of the network.
Assume first that the transition state is fixed in one of the still states
${\bE}\in {\mathcal D}$. In other words, assume that no Long-to-Short and
Short-to-Long transition are allowed.
Let us denote the total average strain as $\be$.
This strain consists of  an eigenstrain  ${\bE}$ and
elastic strain $ \boldsymbol \varepsilon$:
$$ \be={\bE}+\boldsymbol \varepsilon .$$
The network is elastically inhomogeneous. However, we will assume for
simplicity that the elastic properties stay constant. We neglect the
effect of density variation in the transition which is reasonable for
small elongations $s$. Under this assumption, the average elastic energy
of the network is
$$
W(\bvep)=W( \be-{\bE}).
$$
Minimizing the energy over all still state strains ${\bE}$, we conclude
that, for each fixed $n$, the energy density $J_n$ can be written as follows
\begin{equation}
\label{inner}
J_n(\be)= \min_{{\bE}\in{\mathcal D}} W(\be-{\bE})= O(n^{-1}) \quad \If ~ \be \in{\mathcal D}.
\end{equation}
Assume now that $  \be \not \in {\mathcal D}$. The density becomes
\begin{equation}
\label{outer}
J_n(\be)= \min_{{\bE}\in {\mathcal D}} W(\be-\bar{\bE})
\end{equation}

Notice that $J$ is a convex linear combination of the energies of homogeneous
states that correspond to the vertices of ${\mathcal D}$. This follows from convexity of ${\mathcal D}$ and $W(\bvep)$.
Motivated by this, we suggest a representation for the effective energy density $J$ in the limit $n\to\infty$ ($\Gamma$-limit).
By the well known argument \cite{braides_gamma}, using density of piecewise linear deformations, it is enough to consider only linear deformations
(equivalently, constant effective strains). For these strains,
\begin{equation}
\label{eff-energy}
J(\be)=
\left\{
\begin{array}{cc}
\min_{{\bE}_i} W(\be-{\bE}_i), &\quad \If ~ \be \notin{\mathcal D},\\
0 ,\hspace*{2.5cm}&\quad \If ~ \be \in{\mathcal D}.\\
\end{array}
\right.
\end{equation}
Here ${\bE}_i$ are the eigenstrains corresponding to the corners of ${\mathcal D}$.

\section{Conclusions}
We studied the discrete model of solid-solid phase transitions. The model consists of a two-dimensional triangular periodic lattice of nonlinear bi-stable rods.
In order to characterize the effective energy of the lattice, we obtained the description of the set
of the average eigenstrain tensors corresponding to so called still states, e.g. microstructure deformations that
carry no forces. We also proposed several explicit constructions of such states for small and large deformations.

The compatibility conditions for rod elongations were derived. In the small deformation case, these conditions are necessary and sufficient for the existence
of a deformation realizing a given set of elongations.
To characterize the states with zero effective energy, we proposed a construction of a special class of still states.
The eigenstrains of these special
patterns densely cover the set $\mathcal{D}$ of all
eigenstrains corresponding to still states.  We showed that the set
$\mathcal{D}$ for our network is a parallelepiped in 3D space of
independent components of the strain tensor. The orientation and side lengths of ${\mathcal D}$ were explicitly described
in terms of the parameters of the underlying microscopic discrete model: the side lengths are proportional to $s$ and orientation is determined by the components of the lattice direction vectors via the matrix $Q^{-1}$. Components of the inverse matrix $Q$ are defined via components of the direction vectors according to (\ref{eq:matrix_q}).

The above results suggest the following properties of the energy density:
\begin{itemize}
\item if a strain is inside of $\mathcal{D}$, then
the energy density is asymptotically close to zero as the number of the lattice nodes approaches infinity;
\item If the strain is outside of this set,
the energy density is proportional to the square of the distance between the strain and $\mathcal{D}$.
\end{itemize}

\bibliographystyle{siam}
\bibliography{bibliography}

\begin{thebibliography}{10}

\bibitem{braides_interactions}
{\sc N.~Ansini, A.~Braides, and V.~C. Piat}, {\em Interactions between
  homogenization and phase-transition processes}, Tr. Mat. Inst. Steklova, 236
  (2002), pp.~373--385.

\bibitem{braidis}
{\sc N.~Ansini, A.~Braides, and V.~C. Piat}, {\em Gradient theory of phase
  transitions in inhomogeneous media.}, Proc. Roy. Soc. Edin. A, 133 (2003),
  pp.~265--296.

\bibitem{ABP}
{\sc K.~A. Ariyawansa, L.~Berlyand, and A.~Panchenko}, {\em A network model of
  geometrically constrained deformations of granular materials}, Networks and
  Heterogeneous Media, 3 (1) (2008), pp.~125--148.

\bibitem{stress_and_strain_bagi}
{\sc K.~Bagi}, {\em Stress and strain in granular asseblies}, Mechanics of
  Materials, 22 (1996), pp.~165--177.

\bibitem{braides_gamma}
{\sc A.~Braides}, {\em Gamma-convergence for Beginners}, Oxford University
  Press, 2002.

\bibitem{cherkaev_vinogradov}
{\sc A.~Cherkaev, V.~Vinogradov, and S.~Leealavanichkul}, {\em The waves of
  damage in elastic lattices with waiting links}, Design andf Simulation
  Mechanics of Materials, 38 (2006), pp.~748--756.

\bibitem{cherkaev_protective}
{\sc A.~Cherkaev and L.~Zhornitskaya}, {\em Protective structures with waiting
  links and their damage elovulion}, Multibody System Dynamics, 13 (2005),
  pp.~53--67.

\bibitem{theory_elasticity_love}
{\sc A.~E.~H. Love}, {\em A treatise on the mathematical theory of elasticity},
  New York Dover Publications, 1977.

\bibitem{t_definitions_sataki}
{\sc M.~Sataki}, {\em Tensorial form definitions of discrete-mechanical
  quantities for granular assemblies}, International Journal of Solids and
  Structures, 41 (2004), pp.~5775--5791.

\bibitem{cherkaev_1}
{\sc L.~Slepyan, A.~Cherkaev, and E.~Cherkaev}, {\em Transition waves in
  bistable structures. i. delocalization of damage}, Journal of the Mechanics
  and Physics of Solids, 53 (2005), pp.~383--405.

\bibitem{cherkaev_2}
\leavevmode\vrule height 2pt depth -1.6pt width 23pt, {\em Transition waves in
  bistable structures. ii. analytical solution: wave speed and energy
  dissipation}, Journal of the Mechanics and Physics of Solids, 53 (2005),
  pp.~407--436.

\bibitem{slepyan_surprise}
{\sc L.~I. Slepyan and M.~V. Ayzenberg-Stepanenko}, {\em Some surprising
  phenomena in weak-bond fracture of a triangular lattice}, J. Mech. Phys.
  Solids, 50(8) (2002), pp.~1591--1625.

\bibitem{slepyan04-waves}
\leavevmode\vrule height 2pt depth -1.6pt width 23pt, {\em Localized transition
  waves in bistable-bond lattices}, J. Mech. Phys. Solids, 52 (2004),
  pp.~1447--1479.

\bibitem{slepyan-book}
{\sc L.~Splepyan}, {\em Models and Phenomena in Fracture Mechanics},
  Sprnger-Verlag, 2002.

\bibitem{truskinovsky-siam}
{\sc L.~Truskinovsky and A.~Vainchtein}, {\em Kinetics of martensitic phase
  trasformations: lattice model}, SIAM Journ. Appl. Math., 66(2) (2005),
  pp.~533--553.

\bibitem{truskinovsky}
\leavevmode\vrule height 2pt depth -1.6pt width 23pt, {\em Quasicontinuum
  models of dynamic phase transitions}, Continuum Mechanics and Thermodynamics,
  18 (2006), pp.~1--21.

\end{thebibliography}
\end{document}